\documentclass[sigconf,10pt]{acmartaz}

\usepackage{booktabs} 
\usepackage{xspace}
\usepackage{amssymb}
\usepackage{subcaption}

\newcommand{\sparagraph}[1]{\noindent{\bf #1}}
\newcommand{\QPP}{query performance prediction\xspace}
\newcommand{\QEP}{query execution plan\xspace}

\newcommand{\nextSec}{}

\setcopyright{rightsretained}

\acmDOI{}

\acmISBN{XXX-X-XXXXX-XXX-X}

\acmConference[SIGMOD]
\acmYear{2019}

\begin{document}
\title{Plan-Structured  Deep Neural Network  Models for Query Performance Prediction}


\author{Ryan Marcus}
\orcid{0000-0002-1279-1124}
\affiliation{%
  \institution{Brandeis University}
}
\email{ryan@cs.brandeis.edu}
 
\author{ Olga Papaemmanouil}
\affiliation{%
  \institution{Brandeis University}
}
\email{olga@cs.brandeis.edu}

\begin{abstract}

Query performance prediction, the task of predicting the latency of a  query, is one of the most challenging problem in database management systems.  Existing approaches rely on features and performance models engineered by human experts, but often fail to capture the complex interactions between query operators and  input relations, and generally do not adapt naturally to workload characteristics and patterns in query execution plans. In this paper, we argue that \emph{deep learning} can be applied to the \QPP problem, and we introduce a novel neural network architecture for the task: a \emph{plan-structured  neural network}.  Our approach eliminates the need for human-crafted feature selection and automatically discovers complex performance models  both at the operator and query plan level. Our novel neural network architecture can match the structure of any optimizer-selected query execution plan and predict its latency with high accuracy. We also propose a number of optimizations that reduce training overhead without sacrificing effectiveness.  We evaluated our techniques on various workloads and we demonstrate that our plan-structured  neural network can outperform the state-of-the-art in query performance prediction.

\end{abstract}

%
%


\maketitle

\section{Introduction}
Query performance prediction (QPP), the task of predicting the latency of a query, is an important {primitive} for a wide variety of data management tasks, including admission control~\cite{q-cop}, resource management~\cite{step}, and maintaining SLAs~\cite{wisedb-vldb, sla-tree}. QPP is also a notoriously difficult task, as the latency of a query is highly dependent on a number of factors, including, but not limited to, the execution plan chosen and the underlying data distribution. As database management systems become increasing complex, this task only gets more difficult: each new operator or physical design component can introduce new and complex interactions that can be very difficult to model. Optimizer cost models, even when precisely tuned, only attempt to differentiate between potential query execution plans, and serve as poor predictors of query latency on their own~\cite{howgood, opt_est}.

Previous approaches to query performance prediction have focused on designing hand-derived metrics~\cite{contender}, training models based exclusively on plan-level information~\cite{ernest},  proposing mathematical models of relational operators~\cite{LiRobustestimationresource2012}, or combining plan-level and operator-level information in ad-hoc ways~\cite{learning_latency}. All of these methods depend on intelligent human \emph{feature engineering}, the task of selecting or deriving pieces of information from a query plan or query operator that might correlate with its latency. Feature engineering, as a manual process, generally requires significant effort from human experts, but, more importantly, {scales poorly with the increasing complexity of database management systems.}

In this paper, we present a class of deep neural networks (DNNs) capable of performing query performance prediction in a ``human-free'' manner.  DNNs have shown tremendous performance on a number of machine learning tasks~\cite{dnn}, and carry a number of advantages. First, deep neural networks require no human feature engineering beyond the architecture of the network. During training, neural networks \emph{automatically} derive and invent combinations of their inputs that serve as useful features, drastically diminishing the need for human experts~\cite{deep_learning}. Second, DNNs are capable of learning complex models of their training data~\cite{universal_approx}, alleviating the need for ad-hoc and human-derived models of relational operators and their combinatorial interactions.

Despite the proven track record and massive growth of DNNs, applying deep learning to \QPP is not a straight-forward task. DNNs, like many other machine learning algorithms, are designed to map input vectors to  output vectors. However, to act as input to a DNN, query plans need to be carefully vectorized to effectively capture their performance-related properties, such as the tree-based structure of an  execution  plan, features of intermediate results, and all involved operators. Unfortunatelly, existing works applying neural networks to tree structured data~\cite{tree_lstm, tree_conv} are ill-suited for the \QPP task. Specifically, \emph{branch isolation}, the fact that a particular relational operator can only affect the performance of its parents and not its siblings, combined with \emph{heterogeneous tree nodes}, the fact that different  operators have different properties (e.g., number of children, predicates, etc.), make query execution plans a unique structure.

We thus propose a novel deep neural network architecture, a \emph{plan-structured neural network}, specifically crafted for predicting the latency of query execution plans in relational DBMSes. Critically, our structure uses a unique \emph{neural unit}, a small neural network, for each logical operator supported by a DBMS' execution engine. These neural units can model the  latency of an operator  while emitting ``interesting'' features to any subsequent operators in the query plan. These neural units can then be combined together into a tree shape isomorphic to the structure of a given execution plan, creating a single neural network which maps a \QEP directly to a latency. By exploiting \emph{weight sharing}, i.e., the property where the same neural unit is used for any instance of the same operator across plans, these neural units are capable of learning complex interactions between  operators. When compared to simpler models~\cite{LiRobustestimationresource2012,learning_latency,opt_est}, our approach \emph{automatically} models \emph{both} the performance of individual operators \emph{and} the interaction effects between operators, avoiding  the need for costly human model design.

This paper makes the following contributions:
\begin{itemize}
\item{We introduce the notion of {an} \emph{operator-level neural unit,}  a deep neural network that models the behavior of {logical} relational operators. Neural units are designed to produce (a) latency estimates and (b) performance-related properties that might be useful for the latency prediction of their parent operator in the plan.}
\item{We introduce \emph{plan-structured deep neural networks}, a neural network model specifically designed to predict the latency of  query execution plans  by dynamically assembling operator-level neural units in a neural network isomorphic to a given query plan.}
\item{We propose  optimizations for efficiently training our novel neural network architecture, decreasing model training time by nearly an order of magnitude.}
\item{We present experimental results demonstrating that our approach outperforms state-of-the-art techniques for query performance prediction.}
\end{itemize}

{Our work marks the beginning of a new line of research that aggressively uses deep learning to address complex data management problems. While  \QPP has been studied extensively {(e.g., ~\cite{contender,ernest,LiRobustestimationresource2012,learning_latency})}, this is the first approach that completely eliminates human-engineered features and models as well as avoids simplified assumptions made by previous work in attempt to make the \QPP problem tractable by human experts.}


In the next section, we provide background information about deep neural networks. Section~\ref{sec:issues} outlines unique properties of query execution plans that motivate our new deep neural network architecture. Section~\ref{sec:arch} describes our  plan-structured neural network model, and how it can be applied to query execution plans. In Section \ref{sec:model-training}, we describe  critical optimizations to make training a neural network for \QPP tractable. We present experimental results in Section~\ref{sec:experiments}, describe related work in Section~\ref{sec:related}, and offer concluding remarks and directions for future work in Section~\ref{sec:conclusions}.

\nextSec
\section{Neural networks background}\label{sec:nn}
Deep neural networks  (DNNs) have a long-ranging history~\cite{universal_approx}, and have recently enjoyed a surge in popularity~\cite{deep_learning}. This section will cover the basics of DNNs and gradient descent, the primary method for training neural networks. A more detailed discussion can be found in~\cite{deep_survey}.

A DNN model is structured in \emph{layers}, where the first layer takes in a vector representing the input data, and each subsequent layer applies some transformation of the previous layer's output. Each layer consists of nodes which receives input data, {multiplies each input by a coefficient (\emph{weight}), sums the result, and passes that sum through an activation function.}
The activation function introduces non-linearity, allowing neural networks to represent arbitrary functions~\cite{universal_approx}. Intuitively, activation functions helps the network represent the extent that a particular input value affects the ultimate outcome (e.g., ``is this feature helpful is predicting latency data without error?''), and hence determines whether and to what extent that value progresses further through the network to affect the ultimate outcome (i.e., a node's output is ``activated'' or ``deactivated'').

DNNs end in an \emph{output layer}: a {layer}
responsible for mapping the output of the penultimate layer to a prediction. 
DNNs are trained on a dataset, consisting of pairs of \emph{inputs} and \emph{targets}, and aim to learn to accurately map a given input to the correct target. The quality of this mapping is measured by a \emph{loss function}, which quantifies the difference between the neural network's prediction (output) and the ground truth (target). DNNs learn via a process called \emph{gradient descent}, a method that incrementally adjusts the transformation performed by each layer (i.e., the weights), to minimize the loss function. Training a DNN can be seen as a corrective feedback loop, rewarding weights that support correct guesses,  punishing weights that lead to {prediction errors},
and slowly pushing the loss function towards smaller and smaller values. In the process, the network takes advantage of correlations and patterns in the underlying data, creating new, transformed representations of the data. Simultaneously, the network learns to recognize correlations between relevant features and optimal predictions.
Next, we describe the above using more formal definitions, essential for the introduction of our techniques.

\subsection{Layers} 
Each layer  in a DNN  is composed of an affine transformation of its input\footnote{Different application domains may use more complex transformations.} and a non-linear \emph{activation function}. Given an input vector $\vec{x}$ of size $n \times 1$, the $i$-th layer of a network can be defined as a function $t_i(\vec{x})$ that provides an output vector of size $m$:
\begin{equation}
\label{eq:affine}
t_i(\vec{x}) = S\left( W_i \times \vec{x} + \vec{b_i} \right) 
\end{equation}
where $S$ is the activation function and $W_i$ are the \emph{weights} for the $i$-th layer, represented by a matrix of size $m \times n$. The bias, $\vec{b_i}$, is an $m \times 1$ vector representing the constant shift of an affine transform. Together, the weights and the biases represent the \emph{parameters} of the neural network model, and control the properties of the transformation performed at each layer. The activation function, $S$, represents some differentiable but non-liner function: popular choices include a sigmoid function or a rectified linear function~\cite{relu}.

Neural network layers are composed together by feeding the output of one layer into the input of the next. For a neural network with $n$ layers, where $\circ$ represents the {function composition operator}, {a neural network} can be defined as:
\begin{equation}
N(\vec{x}) = t_{n} \circ t_{n-1} \circ \dots t_1
\end{equation}
 For example, a two layer neural network $N$ on an input vector $\vec{x}$ could be represented as:
\begin{equation*}
\begin{split}
N(\vec{x})=t_2(t_1(\vec{x})) =S \left(W_2 \times S\left(W_1 \times \vec{x} + \vec{b_1}\right) + \vec{b_2}\right)
\end{split}
\end{equation*}
\subsection{Gradient descent} \label{sec:gd}

The first step to training a neural network $N$ is defining a \emph{loss function}, i.e. a function whose minimization is a suitable criteria for the network to  provide a good prediction.  Let us assume a  set of training input vectors $X$  and a \emph{labeling function} $l(\vec{x})$ that provides the  corresponding target value for each vector $\vec{x} \in X$. For example, if $X$ is a set of vectors representing query plans, $l(\vec{x})$ could be the latency of the query plan represented by the vector $\vec{x}$. The neural network $N$ can be trained to  produce the  target $l(\vec{x})$ when fed $\vec{x}$ by minimizing a loss function~\cite{sgd}.  One popular loss function is $L_2$ loss, or root mean squared error, which can be defined as: 
\begin{equation}
\label{eq:ex_loss}
err(X) = \sqrt{\frac{1}{|X|} \sum_{\vec{x} \in X} \left( N(\vec{x}) - l(\vec{x}) \right)^2}
\end{equation}
Given a loss function and a dataset, the next task is to adjust the weights and biases of the neural network to minimize the loss function. One popular technique for tweaking the weights and biases  is with gradient descent~\cite{sgd}. Here, the  activation function $S$ is chosen to be differentiable so that the loss function can be differentiated with respect to any particular parameter. This allows us to calculate the \emph{gradient}: the derivative of the loss function with respect to an arbitrary parameter ({i.e. a weight in $W_i$ or a bias in $\vec{b}_i$ for some $i$}
). This represents how the loss function is affected by the particular parameter (i.e., if a higher/lower weight/bias leads to higher/lower loss). The derivative of a particular parameter $w$ (i.e., a single element inside of a weight matrix or bias vector) can be written as the following function:
\begin{equation}
\label{eq:gradient}
\nabla_w(err, X) = \frac{\partial err(X)}{\partial w}
\end{equation}
In other words, the gradient of a parameter of the neural network (weights and biases)  is the \emph{derivative of the loss function with respect to this particular parameter.} This gradient can then be evaluated  given the input vectors $X$ and their corresponding target values $l(\cdot)$.

Gradient descent works by adjusting each weight of the neural network independently. The gradient descent algorithm first computes the  gradient (Equation~\ref{eq:gradient}) of a given neural network parameter, $w$. If the gradient is positive, meaning that an increase in this parameter would (locally) lead to a increase in the loss function, the weight $w$ is decreased. If the gradient is negative, meaning that an increase in this parameter would (locally) lead to a decrease in the loss function, the weight $w$ is increased. After adjusting the weight, the algorithm then repeats this procedure for each parameter in the network. This simple procedure is iterated until gradients of all parameters (weight and biases) are relatively flat (i.e., convergence).

In practice, computing the gradient for the entire dataset (i.e., all of input vectors and target values), is prohibitive. Thus, \emph{stochastic} gradient descent instead takes a simple random sample of the input vectors and their corresponding target values and uses these to \emph{estimate} the gradient~\cite{sgd}.



\nextSec
\section{DL-Based Latency Prediction: Challenges}
\label{sec:issues}

Despite their  numerous advantages, it is difficult to apply traditional deep neural networks to the query performance prediction task.  A straightforward application of deep learning would be to model the whole query as a single neural network and use query plan features as the input vector. However, this naive approach ignores the fact that the query plan structure, features of intermediate results, and non-leaf operator are often correlated with query execution times and hence can be useful in any predictive analysis task. 

Furthermore,  query plans are {diverse} 
structures -- the type and number of operators varies per plan, operators have  different correlations with query performance, and operators have different sets of properties and hence different sets of predictive features.  Traditional DNNs  have static network architectures and deal with input vectors of fixed size. Hence, ``one-size-fits-all'' neural network architectures do not fit the  query performance prediction task. Finally, while previous work in the field of machine learning has examined applying deep neural networks to sequential~\cite{lstm} or tree-structured~\cite{tree_lstm,pollack_ram} data, none of these approaches are ideal for query performance prediction, as we describe next.  

\begin{figure}
\centering
\includegraphics[width=0.40\textwidth]{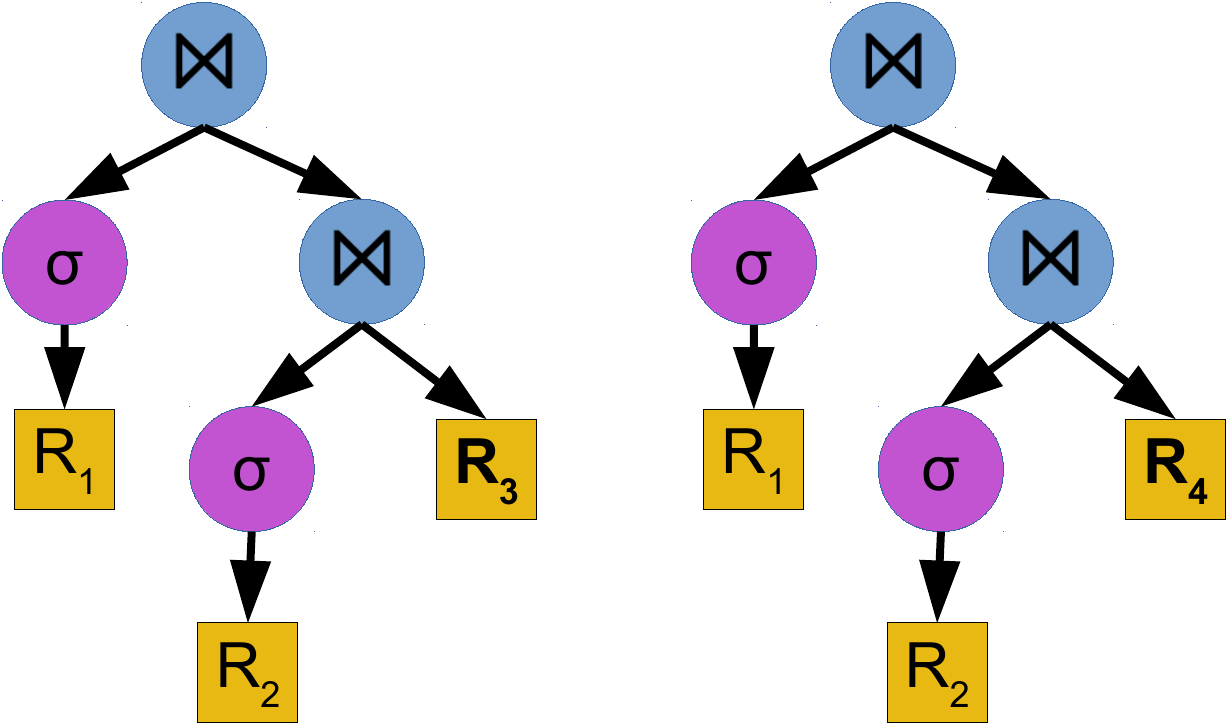}
\caption{\small In query processing, changes in one branch of the query plan \emph{cannot} effect any node outside of their ancestors. Here,  changing the rightmost relation from $R_3$ to $R_4$ will not have any effect on the leftmost filter operation of $R_1$.}
\label{fig:qep}
\end{figure}

\sparagraph{Isolated branches}
Neural network architectures proposed for processing tree-structured data are popular in natural language processing~\cite{tree_lstm,pollack_ram}, and are built around the assumption that a modification to one branch of a tree can have a drastic impact on other branches, allowing  tree branches to share information. However, in the context of a query execution plan, the characteristics and performance of one branch of the query execution plan tree are reasonably isolated from other branches. Specifically, we know that a particular operator can \emph{only} affect its ancestors, and can never affect its siblings. For example, consider the two query execution plans shown in Figure~\ref{fig:qep}. Changing $R_3$ in the first plan to $R_4$ in the second plan cannot affect the performance of $R_1$ or its filter. 

\sparagraph{Heterogeneous tree nodes}
Traditional neural networks operate on input vectors of a fixed structure.
However, in a query execution plan, each type of operator has fundamentally different properties.  A join operator may be described by the join type (e.g. nested loop join, hash join), the estimated required storage (e.g., for an external sort), etc. A filter operation, however, will have an entirely different set of properties, such as selectivity estimation or parallelism flags. Since feature  vectors of different operators are likely of different sizes, simply feeding them into the \emph{same} neural network  is not possible.

A naive solution to this problem might be to concatenate vectors together for each relational operator. For example, if a join operator has 9 properties and a filter operator has 7 properties, one could represent either a join or a filter operator with a vector of size $9+7 = 16$ properties. If the operator is a filter, the first 9 entries of the vector are simply 0, and the last 7 entries of the vector are populated. If the operator is a join, the first 9 entries of the vector are populated and last 7 entries are empty. The problem with this solution is \emph{sparsity}: if one has many different operator types, the vectors used to represent them will have an increasingly larger proportion of zeros. Generally speaking, such sparsity represents a major problem for statistical techniques~\cite{sparse_bad}, and transforming sparse input into usable, dense input is still an active area of research~\cite{VaswaniAttentionAllYou2017,OordNeuralDiscreteRepresentation2017}. In other words, using sparse vectors to overcome heterogeneous tree nodes replaces one problem with a potentially harder problem.

\sparagraph{Position-independent operator behavior} As pointed out by previous work~\cite{LiRobustestimationresource2012, opt_est}, two instances of the same operator (e.g., join, selection, etc), will share similar performance characteristics, even when appearing within different plans or multiple times in the same plan. For example, in the case of a hash join, latency is strongly correlated with the size of the probe and search relations, and this correlation holds \emph{regardless of the operator's position in the query execution plan.} This indicates that one could potentially train a neural network model to predict the performance of a  hash join operator, and that same model can be used any time the hash-join operator appears in a plan.


\nextSec
\section{Plan-structured  DNNs}\label{sec:arch}

 Taking the above observations into account, this paper proposes {a new tree-structured neural network architecture}, in which the structure of the network  matches the structure of a given query plan. This \emph{plan-structured neural network} consists of operator-level neural networks (called \emph{neural units}) and the entire query plan is modeled as a tree of neural units. On its own, each neural unit is expected to (1) predict the performance of an individual operator type  -- for example, the neural unit corresponding to a join predicts the latency of joins -- as well as (2) ``interesting'' data regarding the operator that could be useful to the parent of the neural unit.  The plan-level neural networks is expected to predict the execution time of a given query plan. 
 
Next, we discuss our proposed model in more detail, starting with the operator-level neural units and moving on with the plan-structured neural network architecture. 

\subsection{Operator-level neural units}
\label{sec:units}


Our proposed  approach models each logic operator type supported by a DBMS' execution engine with a unique  neural unit, responsible for learning the performance of that particular operator type, e.g., a unique  unit for joins, a unique  unit for selections, etc. These neural units aim to represent sufficiently complex functions to model the performance of relational operators in a variety of contexts. For example, while a simple polynomial model of a join operator may make predictions only based on estimated input cardinalities, our neural units will automatically  identify the most relevant features out of a wide number of candidate inputs (e.g., underlying structure of the table, statistics about the data distribution, uncertainty in selectivity estimates, available buffer space, etc.), all without any hand-tuning. 

\sparagraph{Input feature vectors} Let us  define as $\vec{x} = F(x)$ a vector representation describing $x$, an instance of a relational operator. This vector will act as an input to the neural unit of that particular operator.  These vectors could be extracted from the output of the query optimizer, and contain information such as: the type of operator (e.g., hash join or nested loop join for join operators, etc.), the estimated number of rows to  be produced, the estimated number of I/Os required, etc. Many DBMSes expose this information through convenient APIs, such as \texttt{EXPLAIN} queries. For examples, see Appendix~\ref{apx:features}, which lists the features used in our experimental study for several operators. Note that the size of the input vector may vary based on its corresponding operator: {input vectors for join operators would have a different set of properties and thus different sizes than the input vectors for selection operators.} However, every \emph{instance} of a relational operator of a given type will have the same size {of input} vector, e.g. all join operators have the same size input vectors.

\sparagraph{Output vector} Performance information for an operator instance $x$  are often relevant to the performance of its parent operator in a \QEP.  To capture this, and allow for flow of information between operator-level neural units, each neural unit of an operator type will output both a latency prediction and a \emph{data vector}. While the latency output predicts the operator's latency, the output data vector  represents  ``interesting'' features from the child operator that are relevant to the performance of the parent operator. For example, a neural unit for a scan operator may produce a data vector that contains information about the expected distribution of the produced rows.  We note these data vectors are learned \emph{automatically} by the model during its training phase, without any human interference or selection of the features that appear in the output vector.


\sparagraph{Neural units} Next, we define a \emph{neural unit} as a neural network $N_A$, with $A$ representing a type of relational operator, e.g. $N_{\bowtie}$ is the neural unit for join operators. For each instance $a$ of the operator type $A$ in a given query plan, the neural unit $N_A$ takes as input the vector representation of the operator instance $a$, $\vec{x_a}$. 

This input is fed through a number of hidden layers, with each hidden layer generating features by applying an activated affine transformation (as defined by Equation~\ref{eq:affine}).  These complex transformations can be learned \emph{automatically} using gradient descent methods, which gradually adjusts the weights and bias of the neural unit $N_A$ in order to minimize its loss function (as described in Section~\ref{sec:gd}). The last layer transforms the internal representation learned by the hidden layers into a latency prediction and an output data vector.

Formally, the output of a neural unit $N_A$ is defined as:
\begin{equation}
\label{eq:neural_unit}
\vec{p_a} = N_{A}\left( \vec{a} \right) \mbox{{, when $a$ is a leaf}}
\end{equation}
\noindent where $a$ is the instance of the operator type $A$. The output vector has a size of $d+1$. The first element of the output vector represents the neural unit's estimation of the operator's latency, denoted as $\vec{p_a}[l]$. The remaining $d$ elements represent the data vector, denoted as $\vec{p_a}[d]$.  We note that since  the input vectors to  different neural units will not have the same size,  each neural unit may have different sizes of weight and bias vectors that define the neural unit, but their fundamental structure will be similar.

\subsubsection{Leaf neural units (scan)} 

The simplest neural units are those representing leaf nodes of the query plan tree and are responsible for accessing data from the database.  Following PostgreSQL terminology, we refer to these as \emph{scan} operators to distinguish them from the selection operator that filters out intermediate data. 

\begin{figure}
\centering
\includegraphics[width=0.38\textwidth]{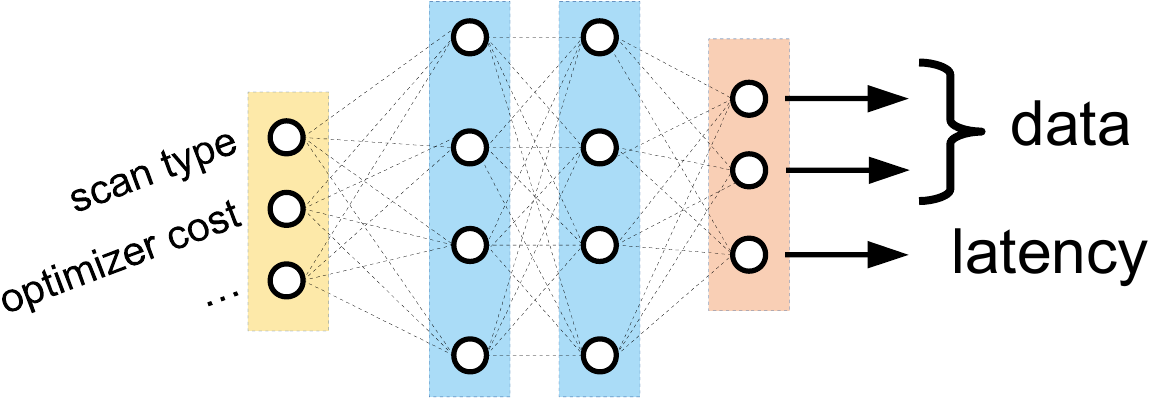}
\vspace{-2mm}
\caption{\small Neural unit corresponding to a scan operator,~$N_{s}$. Input features are mapped through a number of hidden layers and one output layer.}
\label{fig:scan_unit}
\vspace{-2mm}
\end{figure}

For a given instance $s$ of the scan operator type $S$, the neural unit $N_S$ takes as input the raw vector representation of the operator instance $s$, $\vec{x_s}$, and produces an output vector $\vec{p_s}$. Since these operators access {stored} data, their corresponding  vectors include (among others) information about the input relation of the scan operator. These features are collected through the optimizer (or various system calls). We refer to the function collecting this information for an instance $a$ of the operator $A$ as $F(a)$. 

Figure~\ref{fig:scan_unit}  shows an illustration of a neural unit for a scan operator $N_S$. The unit takes information from the query plan (e.g., index/table scan, optimizer's  cost, cardinality estimates,  estimated I/Os, memory availability, etc) as input.  By running the raw vector representation of a scan operator through many successive hidden layers, the neural unit can model complex interactions between the inputs. For example, a series of activated affine transformations might ``trust'' the optimizer's cost model more for certain types of scans, or for scans over particular relations.  The neural unit transforms the input vector  into a latency prediction and a output data vector. Possible output data features here might be  distribution information of the rows emitted by the scan.

\subsubsection{Internal neural units} 
Having constructed neural units for each leaf operator type, we next explain how the internal operators, i.e., operators with children in a \QEP, can be modeled using neural units. Like leaf operators, the neural units for the internal operator instance $x$ of the \QEP will take an operator-specific input vector into account, provided by the function $F(x)$. However, the performance of the internal operators depends also on the behavior of its children. Hence, each internal  neural unit receives also as input the latency prediction and the output data vector of its children.

 A neural unit for an internal operator type $A$ is  represented as a neural network $N_A$. Given a \QEP where an operator instance $a$ of type $A$ receives input from  operators $x_i, i\in \{1,...n\}$, the input vector of $N_A$ will be the operator-related information produced by $F(a)$ in addition to \emph{the output vectors produced by the operator's children}:
\begin{equation}\label{eq:gennu}
\vec{p_{a}} = N_A(F(a) \frown \vec{p_{x_i}} \frown \dots \frown \vec{p_{x_n}})
\end{equation}
\noindent where $\frown$ represents the vector concatenation operator. 

Since the operator's raw input vector and the outputs of its children's neural units are concatenated together and passed as an input to another neural network,  the loss function of the entire neural network architecture is still differentiable with respect to any weight. This guarantees that the network can still be trained using standard gradient descent methods.

\begin{figure}
\centering
\includegraphics[width=0.45\textwidth]{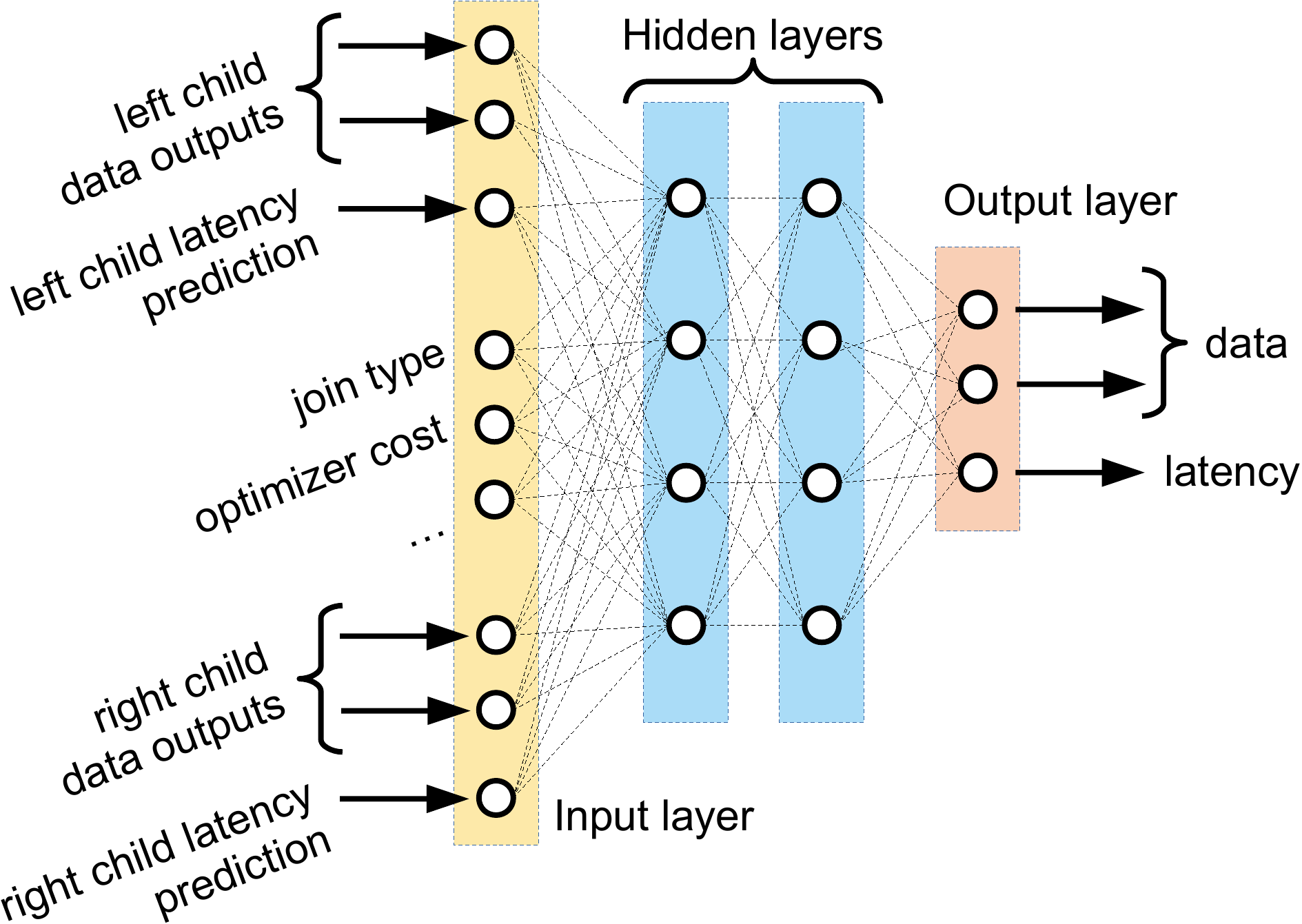}
\caption{\small Neural unit corrosponding to a relational join operator,~$N_{\bowtie}$. The neural unit takes input from two children and several optimizer features.}
\label{fig:nn_detail}
\end{figure}

 Figure~\ref{fig:nn_detail} shows an example of an internal neural unit, corresponding to a join operator, $N_{\bowtie}$. The unit takes information about the join operator itself (e.g., the type of join, the optimizer's predicted cost, cardinality estimates, etc.) as well as information from the join operator's children. Specifically, the join neural unit will receive both the data vector and latency output of its left and right child. These inputs are fed through a number of hidden layers and is transformed into a final output vector, where the first element of the output vector represents the predicted latency and the remaining elements represent the data output features. This allows $N_{\bowtie}$ to be further composed with other neural units. 

\begin{figure}
\centering
\includegraphics[width=0.45\textwidth]{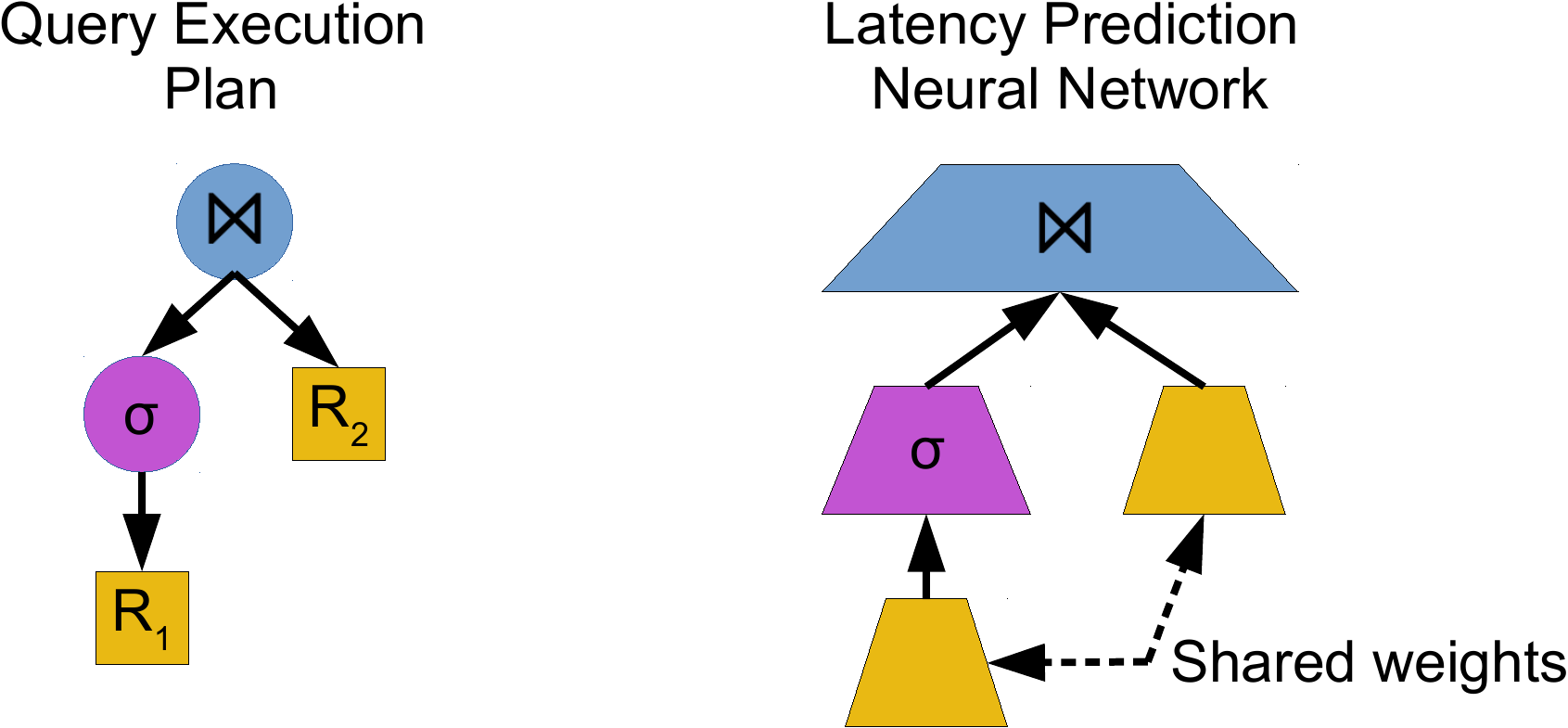}
\caption{\small Each node in a query execution plan is mapped to a neural unit corresponding to the relational operator.}
\label{fig:nn_general}
\end{figure}

\subsection{Trees of neural units}
\label{sec:tree}

Next, we show how these neural units can be composed into tree structures isomorphic to any particular \QEP. Figure~\ref{fig:nn_general} shows an example of a \QEP and the isomorphic plan-structured neural network. Intuitively, each operator in a \QEP is replaced with its corresponding neural unit, e.g.~join operators are replaced with $N_{\bowtie}$, and the output of each neural unit is fed into the parent. The latency of the \QEP is the first element of the output vector $\vec{p_r}$, where $r$ is the instance operator on the root of the \QEP. Note that the recursive definition (Equation~\ref{eq:gennu}) of $\vec{p_r}$ will ``replace'' each relational operator with its corresponding neural unit in a top-down fashion.


\begin{figure}
\centering
\includegraphics[width=0.45\textwidth]{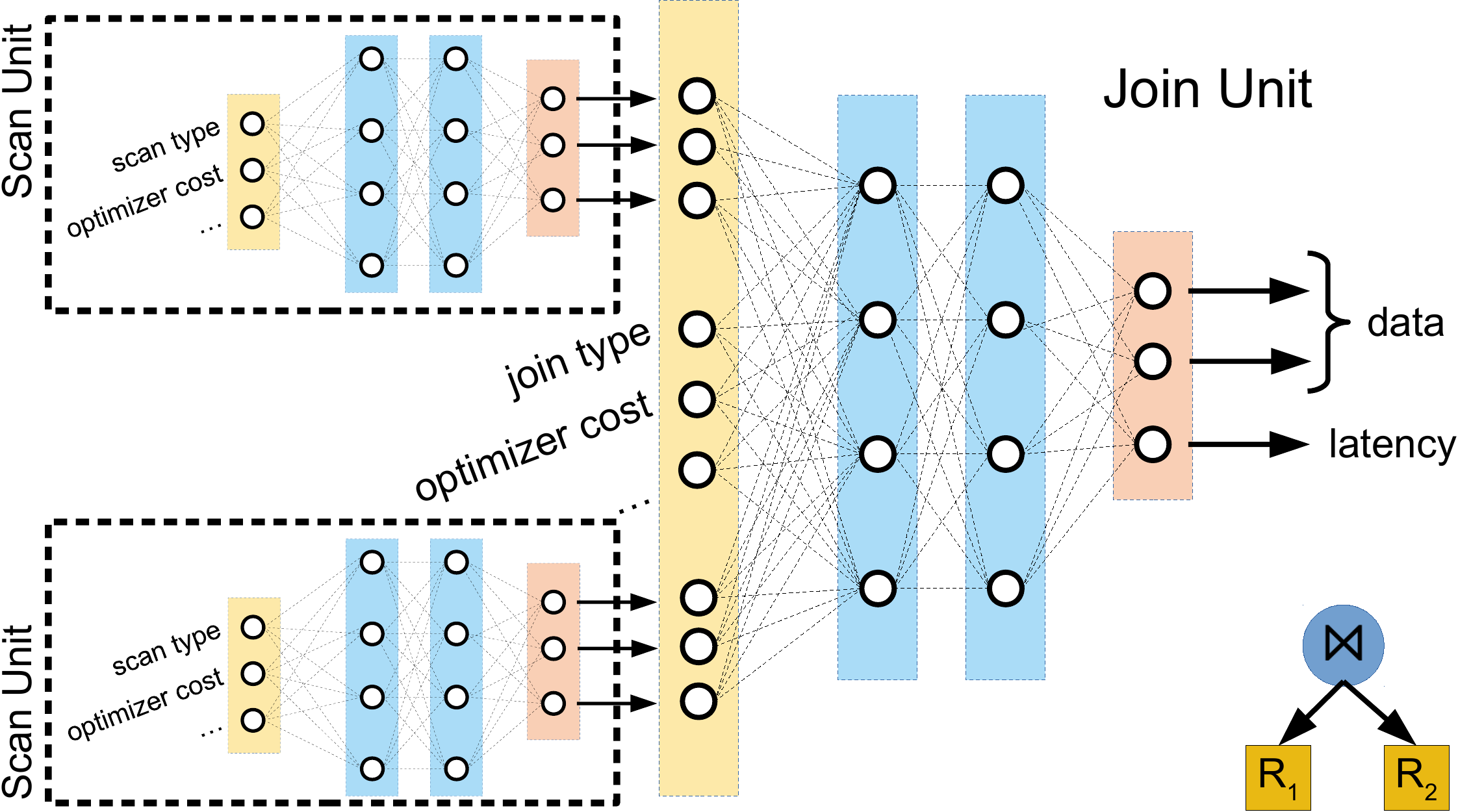}
\caption{\small A neural network for a simple join query}
\label{fig:nn_full}
\end{figure}

Figure~\ref{fig:nn_full} shows an example of this construction. For the \QEP shown in the bottom-right of the figure (two scans and a join), two instances of the scan neural unit and one instance of the join neural unit are created. The outputs of the scan units are concatenated together with information about the join operator to make the input for the join unit, which produces the final latency prediction.

\begin{figure}
\centering
\includegraphics[width=0.34\textwidth]{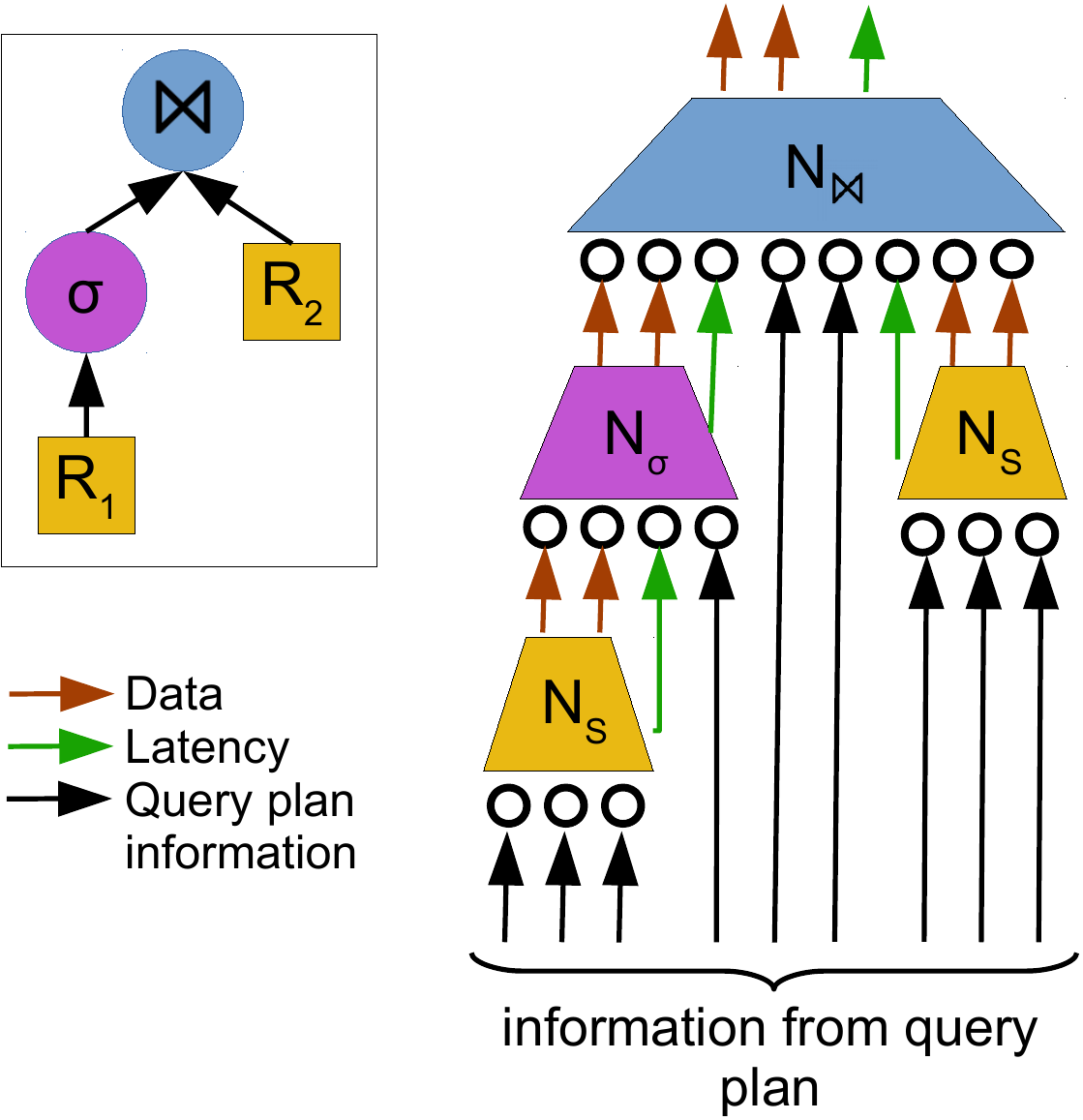}
\vspace{-2mm}
\caption{\small General neural network for latency prediction}
\label{fig:nn_overview}
\vspace{-6mm}
\end{figure}

Figure~\ref{fig:nn_overview} shows a more general example, with a query plan (top left) and the corresponding neural network tree (right). Each neural unit, represented as trapezoids, takes in a number of inputs. For the leaf units (orange, corresponding to the table scans in the query plan), the inputs are information from the query plan (black arrows). The internal, non-leaf units take information from the query plan as well, but additionally take in the latency output (green arrows) and the data outputs (red arrows) of their children. The latency outputs represent the model's estimate of the latency of each operator, and the data outputs contain information about each operator that may be useful to the parent operator (for example, the table scan neural unit may encode data about which relation is being read).  Here, the two orange trapezoids correspond to the scan operators of $R_1$ and $R_2$ in the query plan, and thus use the same neural unit.

\subsection{Model benefits} 

 Our plan-structured neural network model eliminates a number of aforementioned challenges, while leveraging a number of plan-structured properties (see Section~\ref{sec:issues}).

\sparagraph{Branch isolation} Since we know that any particular relational operator in a \QEP can only affect the performance of its ancestors, and not its siblings or children, we say that a \QEP exhibits branch isolation. The way we assemble neural units into trees respects this property: each neural unit passes information exclusively upwards. Intuitively, this upwards-only communication policy directly encodes  knowledge about the structure of the query execution plan into the network architecture itself.

\sparagraph{Heterogeneous tree nodes} Operator-level neural unit accept input vectors of different size depending on the operator they model while producing a fixed-sized output vector. This enables the structure of the plan-structured neural network to dynamically  match any given query plan, thus making our model suitable to handle arbitrary plans. For example, regardless of if the child of a join operator is a filter (selection) or a scan, its child neural unit will produce a vector of a fixed size, allowing this output vector to be  connected to the neural unit representing the join operator.

\sparagraph{Position-independent operator behavior} Since we expect a particular operator to have some common performance characteristics regardless of its position in the \QEP, the same neural unit is used for every instance of a particular operator. Because the same \QEP can contain multiple instance of the same operator type (e.g., multiple joins), our architecture can be considered a \emph{recurrent neural network}~\cite{rnn}, and as such benefits from \emph{weight sharing}: since instances of the same operators share similar properties, representing them with a single neural unit (and thus a a single set of weights and bias) is both efficient and effective. However, since \emph{distinct} operator types are represented by different neural units (and thus will \emph{not} share the same weights and bias), our approach can handle the heterogenous nature of the query execution plan operators.



\nextSec
\section{Model training}\label{sec:model-training}

So far, we have discussed how to assemble neural units into trees matching the structure of a given \QEP. In this section, we describe how these plan-structured neural networks are trained. Training is the process of progressively refining the network's weights and bias (in our case, the weight and bias of the neural units included in a plan-structured neural network) to minimize the loss function using gradient descent (as discussed in  Section~\ref{sec:nn}). 


Initially, each hidden layer, and the final latency and data vector output for each neural operator, will simply be a random activated affine transformation. In other words, the weights and bias that define the transformations (Equation~\ref{eq:affine}) are initially picked randomly. Through repeated applications of gradient descent, these transformations are slowly tweaked to map their inputs slightly closer to the desired target outputs.  

This training process is performed using a large corpus of executed query plans. Formally, for a dataset of executed query execution plans, let $D$ be the set of all query operator instances within those plans. Then, for each query operator $o \in D$, let $l(o)$ be the latency of the operator. The neural units are trained by minimizing the following loss function:


\begin{equation}
\label{eq:loss}
L_2(D) = \sqrt{\frac{1}{|D|} \sum_{o \in D} \left(\vec{p_o}[l] - l(o)\right)^2}
\end{equation}
\noindent where $\vec{p_o}[l]$ represents the latency output of the operator $o$'s neural unit (the neural unit's prediction).


Note that, if a particular query operator instance $o$ is not a leaf in its \QEP, the evaluation of its output $\vec{p_o}$ will involve multiple neural units, based on the recursive definition given by Equation~\ref{eq:gennu}. The loss function $L_2(D)$ thus represents the prediction accuracy of leaf operators, internal operators, \emph{and} the root operator of a \QEP. Minimizing this loss function thus minimizes the prediction error for \emph{all} the operators in the training set.

Intuitively, Equation~\ref{eq:loss} is simply the combined differences between the latency the model predicted for each operator and the observed, ``ground truth'' latency. The loss function explicitly compares the predicted latency of a particular operator $\vec{p_o}[l]$ with the ``ground truth'' latency $l(op)$.

However, it is important to note that the loss function does \emph{not} explicitly compare the  data vector of the output $\vec{p_o}$ to any particular value. The gradient descent algorithm is thus free to tweak the transformations creating the output data vector  to produce useful information for the parent neural unit consuming the data vector output. To exemplify this, consider evaluating the output $\vec{p_j}$ for a simple query plan involving the join $j$ of two relation scans, $s_1$ and $s_2$, as illustrated in Figure~\ref{fig:nn_full}. Following Equation~\ref{eq:gennu}, and defining as $N_{\bowtie}$ the neural unit for joins and $N_S$ the neural unit for scans, we can expand the output vector $\vec{p_j}$ as follows:
\begin{align*}
  \vec{p_j} & = N_{\bowtie}\Big(F(j) \frown \vec{p_{s_1}}  \frown \vec{p_{s_2}}  )\\
       & = N_{\bowtie}\Big(F(j) \frown N_S(F(s_1)) \frown N_S(F(s_2)) \Big) \\
       & = N_{\bowtie}\Big(F(j) \frown [\vec{p_{s_1}}[l] \frown \vec{p_{s_1}}[d]] \frown [\vec{p_{s_2}}[l]  \frown \vec{p_{s_2}}[d]]\Big) 
  \end{align*}
\noindent where $\vec{p_{s}}[d]$ represents the output data  vector for the neural unit of operator $s$.

Thus, the transformations producing the output data vector  for each neural unit are adjusted by the gradient descent algorithm to minimize the latency prediction error of their \emph{parents}. In this way, each neural unit can learn what information about its represented operator type is relevant to the performance of the parent operator \emph{automatically}, without expert human analysis. Because the training process does not push output data vectors to represent any pre-specified values, we refer to these values as \emph{opaque}, as the exact semantics of the output data vector  will likely vary significantly based on context, and may be difficult to interpret directly, as is generally the case with recurrent neural networks~\cite{rnn}.

\subsection{Training optimizations} 

The training overhead of our plan-structured neural network model can be significant due to the large number of operators that might appear in a \QEP, and thus we wish to train multiple neural units in parallel. To address this challenge, we  propose two optimizations over naive training methods. These techniques aim to improve the \emph{performance of} computing of the loss function of a plan-structured neural network (Equation~\ref{eq:loss}) in the context of gradient descent. Section~\ref{sec:batch} explains how the loss function can be computed efficiently in a vectorized way. Section~\ref{sec:shared-info} notes that computing the loss function can be accelerated by exploiting the tree structure of a \QEP. 


\subsubsection{Batch training}\label{sec:batch}
Gradient descent minimizes a neural network's loss function by tweaking each weight by a small amount based on the gradient of that weight (Equation~\ref{eq:gradient}). Ideally, the gradient of each weight should be evaluated over the whole training set.  However, since neural networks are trained over very large datasets, naive implementations of gradient descent are \emph{space} prohibitive. Furthermore, since real-world neural networks can have a significant number of weights, naive implementations of gradient descent which update each weight sequentially are \emph{time} prohibitive.

To reduce space usage, modern differentiable programming frameworks~\cite{tensorflow, pytorch} preform training in \emph{batches}. Instead of computing the gradient using the entire dataset (which cannot effectively fit into memory), simple random samples (called batches or mini-batches) are drawn from the data and used to estimate the gradient and adjust the weights. This widely-adopted technique is called \emph{stochastic gradient descent}~\cite{sgd}. Since each sample is selected at random, the estimation of the gradient is unbiased~\cite{BottouLargeScaleMachineLearning2010}.

To reduce the time required, modern neural network libraries take advantage of vectorization (i.e., {applying mathematical operators to entire vectors simultaneously}) to speed up their models. To do so, neural networks are assumed to be using a fixed architecture, e.g. an architecture that does not change based on the particular input. This is the case in many applications, such as computer vision. {By assuming a fixed architecture, libraries can assume that the {symbolic} gradient of each weight will be identical for each item in the batch {(i.e., the derivative of any weight can be computed using the same sequence of mathematical operations)}, and thus their computation can be vectorized.}

%
%
%


Stochastic gradient descent and vectorization work very well for neural networks where the structure of the network does not change based on the inputs.  However, these optimization poses a challenge for our proposed plan-structured neural network model: if two samples (i.e., query plan) in a batch have different tree structures, the {symbolic derivative} for a given weight will be vary depending on the input sample, {and thus the sequence of mathematical operations needed to compute the derivative of two given weights can differ. Thus, vectorization cannot be directly applied.}

One solution might be to group the training set into query execution plans with identical structure, and then use each group as a batch. While this would both reduce memory requirements and allow for standard vectorization approaches to be used, most of the effectiveness of stochastic gradient descent depends on the batch being a true simple random sample, and thus providing an unbiased estimate of the gradient~\cite{BottouLargeScaleMachineLearning2010}. By only creating training batches with identical query plans, each batch, and thus each estimation of the gradient, will become biased.

\sparagraph{Plan-based batch training} To address these challenges, we proposes constructing large batches of randomly sampled query plans. Within each large batch $B$, we group together sets of query plans with identical structure, i.e. we partition $B$ into equivalence classes $c_1, c_2, \dots, c_n$ based on plan's tree structure, such that $\bigcup_{i=1}^n c_i = B$. Then, we estimate the gradient as follows:
\begin{equation*}
\nabla_w(L_2, B) = \frac{1}{\sum_{i=1}^n |c_i|} \times \sum_{i = 1}^n \left( |c_i| \frac{\partial L_2(c_i)}{\partial w} \right)
\end{equation*}
The gradient of each weight for each operator is then efficiently estimated within each inner-batch group, and the results are added together and normalized based on group size, ensuring the estimate of the gradient is not biased.

\noindent\textit{Example}: For example, if a randomly-sampled large batch contained three groups of query execution plans with distinct tree structures, $B = \{c_1 \cup c_2 \cup c_3\}$, with 10, 20, and 300 members each, the gradient of each weight would be estimated as follows:
\begin{align*}
  &\nabla_w (L_2, B) = \frac{1}{10+20+300} \times \\
  &\bigg( 10 \frac{\partial L_2(c_1)}{\partial w} + 20 \frac{\partial L_2(c_2)}{\partial w} + 300 \frac{\partial L_2(c_3)}{\partial w}\bigg)\\
\end{align*}
This approach avoids biasing the gradient estimate while still gaining efficiency from batch processing. 

\subsubsection{Information sharing in subtrees}\label{sec:shared-info}

Let us assume that $r$ of type $R$ is the root operator of a \QEP, and $c$ is the sole child of that operator. When computing the loss function of the \QEP's neural network, estimating the latency prediction error of the root, $(\vec{p_r}[l] - l(r))$ in Equation~\ref{eq:loss}, requires us to compute the output vector of its child $c$, $\vec{p_c}$, as an intermediate value. This follows from the definition of $\vec{p_r}$ in Equation~\ref{eq:gennu}, based on which  $\vec{p_r} = N_R(F(r) \frown \vec{p_c})$. Since computing the loss function will \emph{also} require computing $\vec{p_c}$, i.e. in the term $(\vec{p_c} - l(c))$, we can avoid a significant redundant computation by caching the value of $\vec{p_c}$ so that it only needs to be computed once. 

More generally, for an arbitrary root $r \in D$ of a \QEP, we can compute the error of each neural unit in the plan's neural network rooted at $r$ in a bottom-up fashion: first, for each leaf node $leaf$ in the tree rooted at $r$, compute and store the output the neural units corresponding to the leaf nodes, $\vec{p_{leaf}}$. Then, compute and sum the $(\vec{p_{leaf}}[l] - l(leaf))^2$ values, storing the result into a global accumulator variable. Once all the leaf nodes have been resolved in this way, repeat the process moving one level up the tree. When the root of the tree has been reached, the global accumulator value will contain the $(\vec{p_x}[l] - l(x))^2$ values for every operator $x$ in the tree rooted at $r$. The global accumulator  contains the sum of the squared differences between the predicted latency and the actual latency for every node in the tree. Applying this technique over every plan-structured neural network in $D$ can greatly accelerate the computation of our loss function as defined in Equation~\ref{eq:loss}.



%

\nextSec
\section{Experimental Results}\label{sec:experiments}
In this section, we describe the experimental study we conducted for our proposed plan-based neural network model. In all our experiments, our queries were executed on PostgreSQL~\cite{url-postgres} on a single node with an Intel Xeon CPU E5-2640 v4 processor, 32GB of RAM, and a solid-state drive.

\sparagraph{Workload} We conducted experiments using  TPC-H~\cite{tpch}, a decision support benchmark,  and TPC-DS~\cite{tpcds}, a decision support benchmark with a focus on higher volumes of data and more complex queries. All TPC-H query templates were used but only seventy (70) TPC-DS query templates are compatible with PostgreSQL (without modification), hence we use only these templates for TPC-DS.  For both benchmarks, 20,000 queries were executed with a scale factor of 100GB. 

Each query was executed from a ``cold cache'' state (both the OS and PostgreSQL cache) in isolation (no multiprocessing). Execution times and execution plans were recorded using PostgreSQL's \texttt{EXPLAIN ANALYZE} capability. The input features used for each neural unit are those that PostgreSQL makes available through the \texttt{EXPLAIN} command before a query is executed. See Appendix~\ref{apx:features} for a listing.

\sparagraph{Training data} The queries was split into a training set and a testing set in two different ways. For the TPC-DS queries, all of the instances of 10 randomly selected query templates are ``held out'' of the training set (e.g., the neural network trains on 60 query templates, and the performance of the network is measured on instances of the unseen 10 query templates). For the TPC-H data, since there are not enough query templates to use the same strategy, 10\% of the queries, selected at random, are ``held out'' of the training set (e.g., the neural network trains on 90\% of the data, and the performance of the network is measured on the other 10\%).

\sparagraph{Neural networks} Unless otherwise stated, each neural unit had 5 hidden layers, each with 128 neurons each. The data output size was set to $d = 32$. Rectified linear units (ReLUs~\cite{relu}) were used as activation functions. Standard stochastic gradient descent (SGD) was used to train the network, with a learning rate of 0.001 and a momentum of 0.9. Training was conducted over 1000 epochs (full passes over the training queries), which consistently produced the reported results. We used the PyTorch~\cite{pytorch} library to implement the neural network, and we used its built-in SGD implementation.

\sparagraph{Evaluation techniques} We compare our plan-based neural network model (\texttt{QPP Net}) with three other latency prediction approaches: 
\begin{enumerate}
\item{\emph{SVM-based models (\texttt{SVM})}: We implemented the learning-based approach proposed in~\cite{learning_latency}, the state-of-the art in latency prediction for relational queries with no explicit human modeling. Here, a regression variant of SVM (Support Vector Machine) models are built for each operator while selective applications of plan-level models are used in situations where the operator-level models are likely to be inaccurate. In contrast to our approach, the set of input vectors for both the operator and plan level models are hand-picked through extensive experimentation.}
  
\item{\emph{Resource-Based Features (\texttt{RBF})}: We also implemented a predictive model that take as input the features proposed by~\cite{LiRobustestimationresource2012}. Although these features are picked for predicting resource utilization of query operators and by extension query plans, resource usage can be good indicator of query performance in non-concurrent query executions.  Hence we modified the MART regression trees used in~\cite{LiRobustestimationresource2012} to predict query latency. Similarly to the \texttt{SVM} approach, the input features of this model are hand-picked and not automatically engineered as in \texttt{QPP Net}. {However, unlike the \texttt{SVM} approach, the \texttt{RBF} approach uses human-derived models for capturing operator interactions.}}

\item{\emph{Tuned Analytic Model (\texttt{TAM})}}: We also implemented a version of the tuned optimizer cost model model proposed in~\cite{opt_est}. This approach uses the optimizer cost model estimate to predict query latency. First, some ``calibration queries'' are ran to determine the coefficients for the ``calibrated cost model.'' Then, this calibrated cost model is used to predict the query latency using the optimizer's cardinality estimates as inputs.\footnote{For consistency, our version of~\cite{opt_est} uses optimizer estimates of cardinalities as inputs without the proposed ``data sampling'' optimization.} {The \texttt{TAM} approach is thus entirely human-engineered, except for a sparse number of tuned parameters that are adjusted using special calibration queries.}



\end{enumerate}


\sparagraph{Evaluation metrics} To evaluate the prediction accuracy of these techniques, we use two metrics: \emph{relative prediction error} and \emph{mean absolute prediction error}. The relative prediction error has been used in~\cite{learning_latency,LiRobustestimationresource2012} and can be defined as follows. Letting $Q$ be the set of test queries, letting $predicted(q \in Q)$ be the predicted latency of $q$, and letting $actual(q \in Q)$ be the actual latency, the relative prediction error is:

\begin{equation*}
\frac{1}{|Q|} \sum_{q \in Q} \frac{|actual(q) - predicted(q)|}{actual(q)}
\end{equation*}

 However, the ``relative error'' metric has several known flaws~\cite{no_percent_error}. Specifically, relative error systematically favors underestimates due to the asymmetry in the error function. No matter how bad an under-prediction is, the worst value the relative error can take on is 0. However, for over-predictions, the relative error is unbounded, hence the asymmetry. 
Because of the issues with relative error, we also report the mean absolute error, a standard metric for evaluating regression~\cite{scikit-learn}, which symmetrically penalizes under and over estimations:
\begin{equation*}
 \frac{1}{|Q|} \sum_{q \in Q} |actual(q) - predicted(q)|
\end{equation*}
A useful property of mean absolute error is that it shares the same units as the regression target. In our case, since we are predicting a quantity in units of time, the units of the mean absolute error are also time units.

We also report $R(q)$, the maximum of the ratio between the actual and the predicted and the ratio between the predicted and the actual (not to be confused with the coefficient of determination):
\begin{equation*}
R(q) = \max \left( \frac{actual(q)}{predicted(q)}, \frac{predicted(q)}{actual(q)} \right)
\end{equation*}

Intuitively, the $R(q)$ value represents the ``factor'' by which a particular estimate was off. For example, if a model estimates a query's $q$ latency to be 2 minutes, but the latency of the query is actually 1 minute, the $R(q)$ value would be 2, as the model was off by a factor of two. Similarly, if the model estimates a query's latency to be 2 minutes, but the latency of the query is actually 4 minutes, the $R(q)$ value would also be 2, as the model was again off by a factor of two.

\begin{figure}
  \centering
  \begin{subfigure}{0.48\textwidth}
    \includegraphics[width=\textwidth]{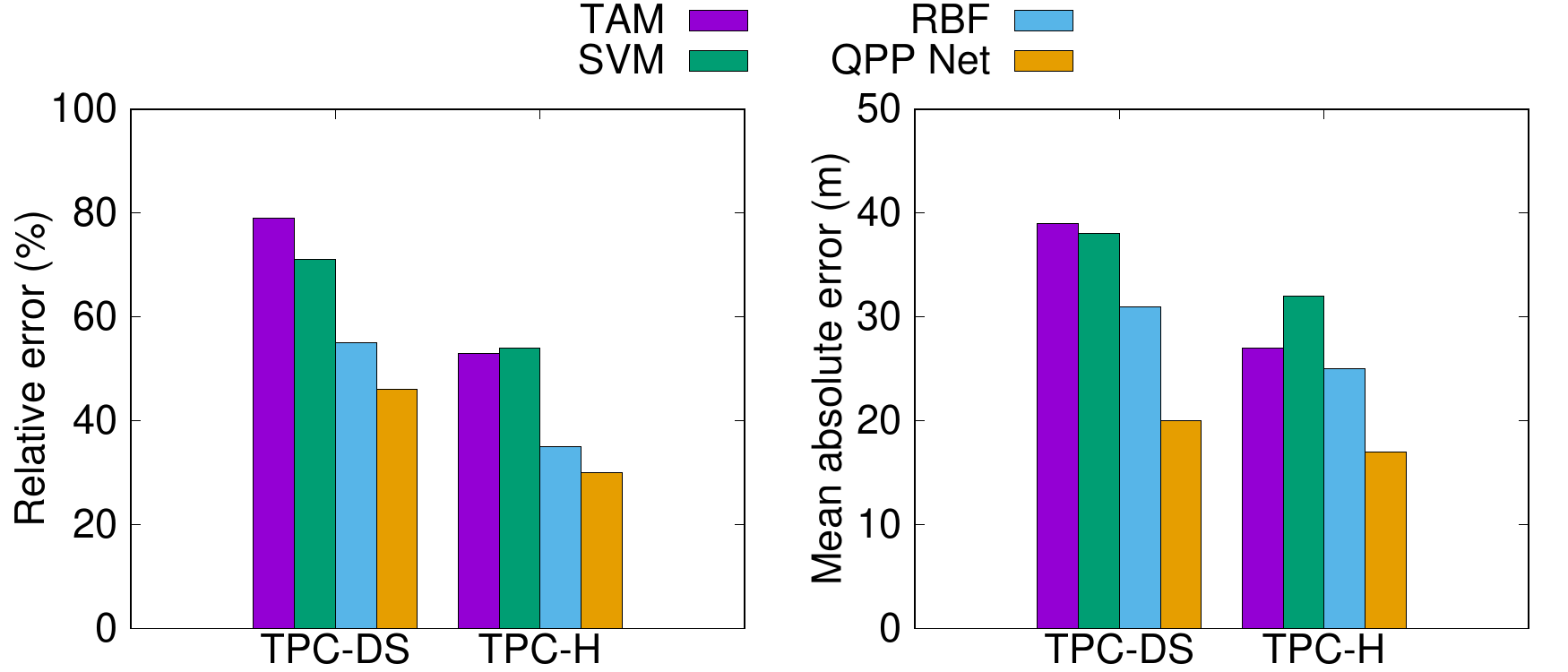}
    \caption{\small Relative error and mean absolute error (lower is better)}
    \label{fig:nn_results}
  \end{subfigure}
  \begin{subfigure}{0.48\textwidth}
    \centering
    \includegraphics[width=\textwidth]{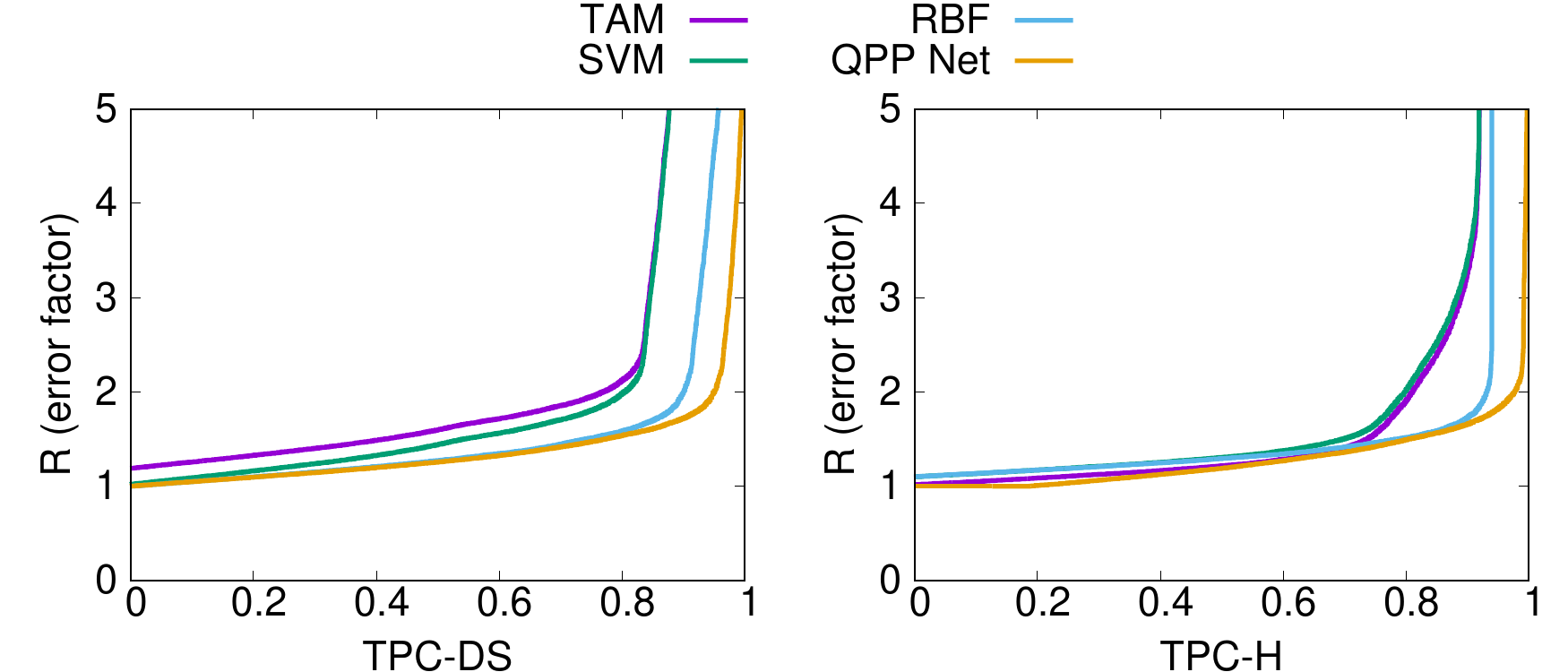}
    \caption{\small Cumulative error factors on both datasets. The x-axis signifies the proportion of the test set for which the algorithm achieved an $R$ score below the value on the y-axis.}
    \label{fig:nn_results_cdf}
  \end{subfigure}
  \label{fig:results}
  \vspace{-2mm}
  \caption{Comparison of prediction accuracy}
  \vspace{-3mm}
\end{figure}

\subsection{Prediction Accuracy}
The accuracy of each method at estimating the latency of queries in the TPC-H and TPC-DS workloads are shown in Figure~\ref{fig:nn_results}. The results reveal that out neural network approach outperforms the other baselines. The relative error improved by 9\% (TPC-DS) and 5\% (TPC-H) over \texttt{RBF}, by 25\% (TPC-DS) and 24\% (TPC-H) over \texttt{SVM} and {by 28\% (TPC-DS) and 21\% (TPC-H) over \texttt{TAM}}. In terms of absolute error, the average error decreased by 11 minutes (TPC-DS) and 7 minutes (TPC-H) from \texttt{RBF}, and by 18 minutes (TPC-DS) and 15 minutes (TPC-H) from \texttt{SVM} and {21 minutes (TPC-DS) and 13 minutes (TPC-H) from \texttt{TAM}}.

The larger improvement seen in TPC-DS is  due to two factors: (1) the fact that the average TPC-DS query plan has more operators in it than the average TPC-H query plan (22 operators vs 18 operators), resulting in \texttt{QPP Net}  being able to take advantage of a larger amount of training data, and (2), \texttt{QPP Net} is capable of learning the more complex interactions present in the TPC-DS workload effectively. \emph{Overall, the hand-picked features used by the other techniques fail to capture the complex interaction between operators, while our automatic features selection approach outperforms all human-crafted techniques, with more significant gains when the query workload is more complex.}

\begin{table}[]
  \begin{subtable}[]{0.48\textwidth}
    \centering
\begin{tabular}{ccccc}
\toprule
Model           & $R \leq 1.5$ & $1.5 < R < 2.0$ & $2.0 \geq R$   \\ \midrule
QPP Net         &  89\%  & 7\%  &  4\%   \\
TAM             &  51\%  & 22\% &  27\%   \\
SVM             &  68\%  & 15\%  & 17\%  \\
RBF             &  85\%  & 6\%  & 9\%   \\
\end{tabular}
\caption{TPC-DS}
\label{tbl:factor_ds}
\end{subtable}
\begin{subtable}[]{0.48\textwidth}
  \centering
  \begin{tabular}{ccccc}
\toprule
Model           & $R \leq 1.5$ & $1.5 < R < 2.0$ & $2.0 \geq R$   \\ \midrule
QPP Net         &  93\%  & 6\%  & 1\%   \\
TAM             &  78\%  & 17\% & 5\%   \\ 
SVM             &  72\%  & 20\%  & 8\%  \\
RBF             &  88\%  & 6\%  & 6\%   \\
\end{tabular}
\caption{TPC-H}
\label{tbl:factor_h}
\end{subtable}
\caption{Percentage of the test set where a particular model's estimate was within a factor of 1.5 of the correct latency, within some factor between 1.5 and 2, and not within a factor of 2.}
\vspace{-12mm}
\end{table}

\subsubsection{Prediction distribution}
We analyzed how frequently each model's prediction was within a certain relative range of the correct latency. We report the percentage of the test set for which each model's prediction was closer than a factor of 1.5 from the actual latency, between a factor of 1.5 and 2 of the actual latency, and greater than a factor of 2 from the actual latency. Tables~\ref{tbl:factor_ds} and~\ref{tbl:factor_h} display the results for TPC-DS and TPC-H, respectively.

For both workloads, \texttt{QPP Net} has the best performance in terms of the proportion of the test set with an error factor less than 1.5. A high percentage of its predictions (89\% for TPC-DS and 93\% for TCP-H) are only within a factor of 1.5 of the actual latency, outperforming \texttt{TAM} by 38\% (TPC-DS) and 15\% (TPC-H), \texttt{SVM} by 21\% (both TPC-H and TPC-DS) and \texttt{RBF} by 4\% (TPC-DS) and 5\% (TPC-H). The results indicate that our approach offers predictions closer to the real latency for a significantly higher number of queries. 


We additionally plot the distribution of $R(g)$ values in Figure~\ref{fig:nn_results_cdf}, in the style of cumulative density function. Each plot shows the largest $R(g)$ value achieved for a given percentage of the test set. For example, on the left hand graph for the QPP Net line, at $0.93$ on the x-axis, the y-axis value is ``1.5''. This signifies that QPP Net's prediction was within at least a factor of 1.5 for 93\% of the testing data. For both datasets, QPP Net's curve has a smaller slope, and does not spike until it is much closer to 1 than the other curves. This means the \emph{our estimates are within a lower error factor for a larger portion of the testing queries compared with the other techniques.}

\subsubsection{Errors by query template}
\begin{figure*}
\centering
\includegraphics[width=\textwidth]{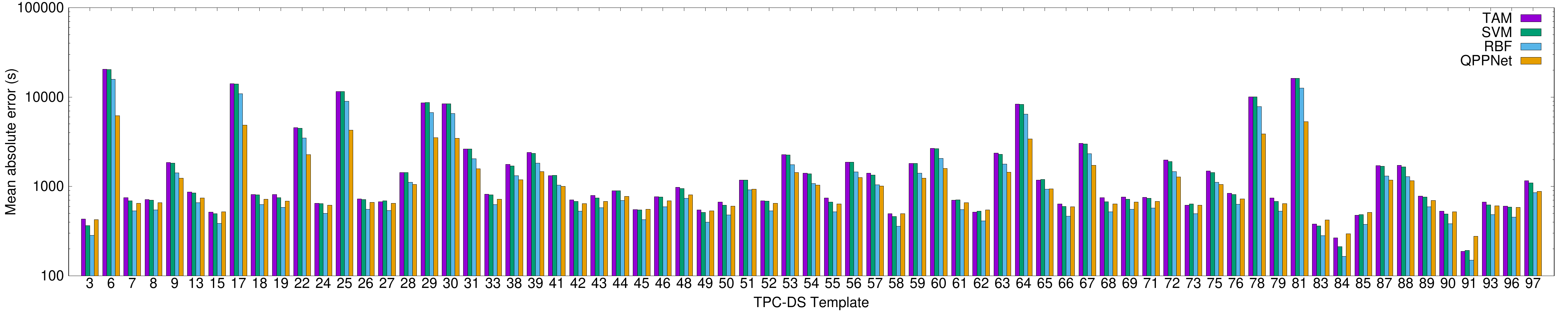}
\caption{ Mean absolute error by template for TPC-DS}
\label{fig:by_template_ds}
\end{figure*}

Figure~\ref{fig:by_template_ds} shows the mean absolute error of the queries in the test set grouped by the TPC-DS query template (note the log scale). {For this experiment, we use a ``hold one out'' strategy: the model is trained using all but one of the query templates, and then the performance on the held-out query template is measured.}

For each query template, the mean absolute error of QPP Net is either lower or within 5\% of the other models. {Results for TPC-H are similar, but as TPC-H has far fewer templates, we omit the graph.} We therefore conclude that \emph{QPP Net can accurately predict the latency of queries across a wide variety of query templates.}

Generally, QPP Net's performance greatly exceeds the other models on the query templates with the highest average latency (e.g. TPC-DS templates 6, 17, 81). A plot of the mean latency of each query template is provided in Appendix~\ref{apx:latency}. While the mean absolute error on these templates exceeds that of the other templates, the relative performance of QPP Net compared to the other models increases significantly. Thus, \emph{compared with other models, QPP Net performs especially well on long-running queries}.


\subsection{Training Overhead}
In this section, we evaluate various properties and behaviors of the neural network model during training, including analyzing the effectiveness of our proposed optimizations.

\begin{figure*}
  \centering
  \begin{subfigure}{0.33\textwidth}
    \includegraphics[width=\textwidth]{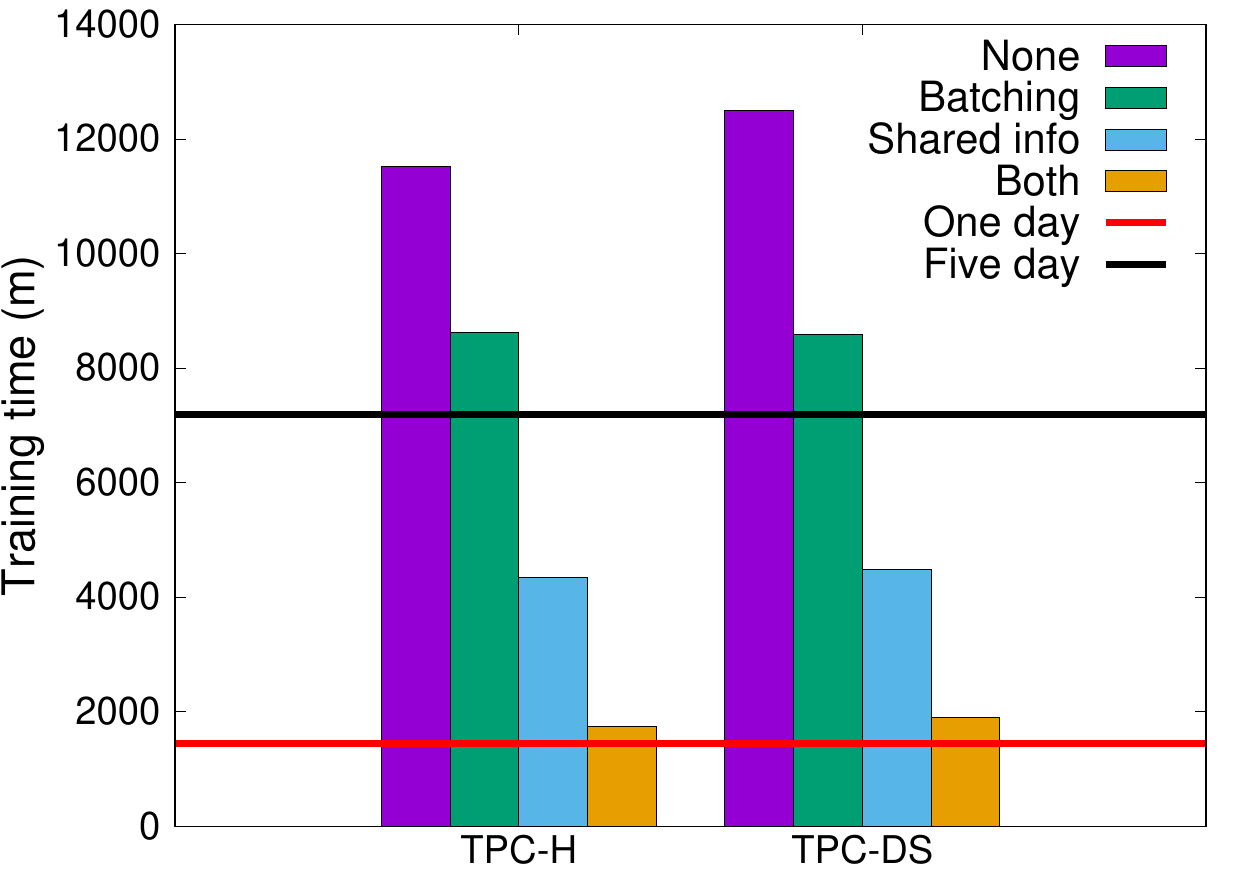}
    \caption{\small Impact of training optimizations}
    \label{fig:train_opts}
  \end{subfigure}
  \begin{subfigure}{0.33\textwidth}
    \centering
    \includegraphics[width=\textwidth]{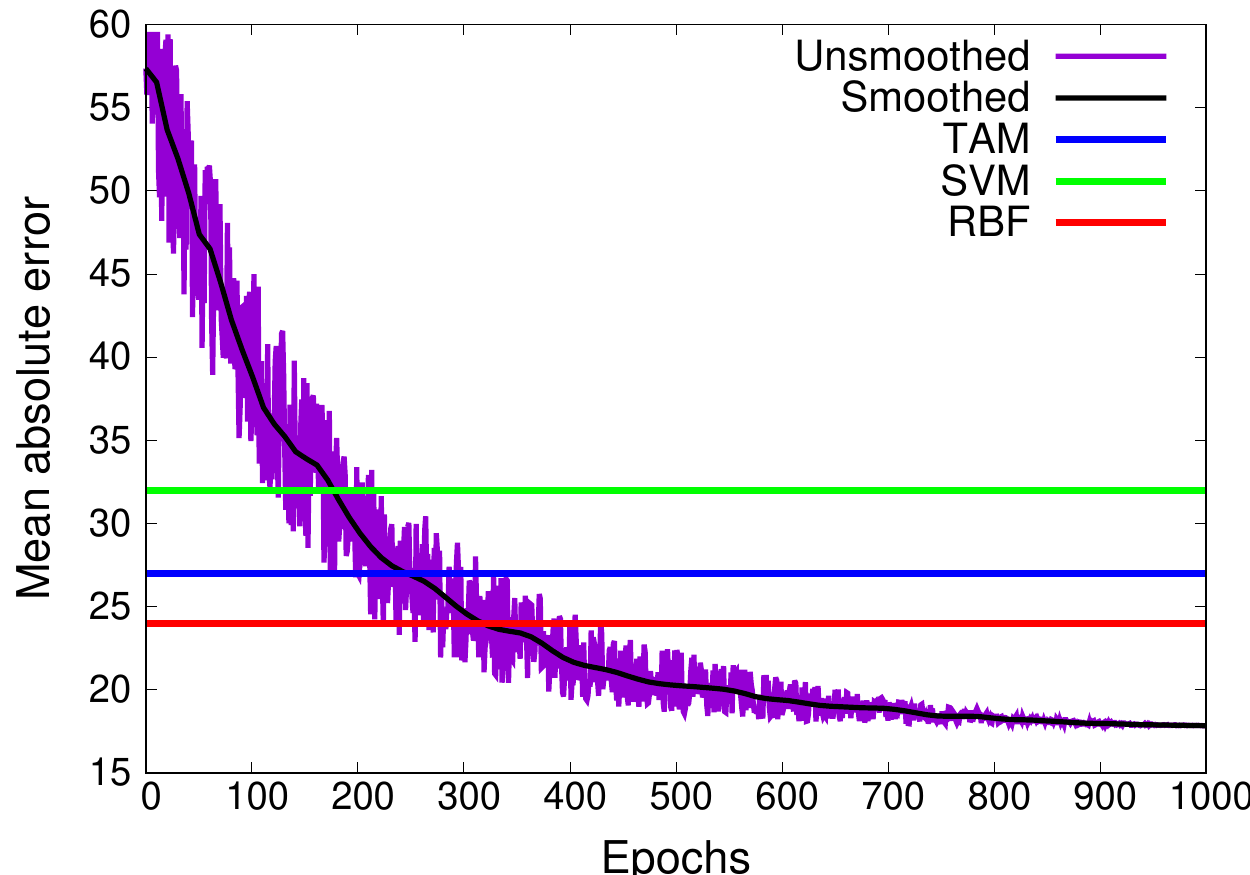}
    \caption{\small Training convergence (TPC-H)}
    \label{fig:train_h}
  \end{subfigure}
  \begin{subfigure}{0.33\textwidth}
    \centering
    \includegraphics[width=\textwidth]{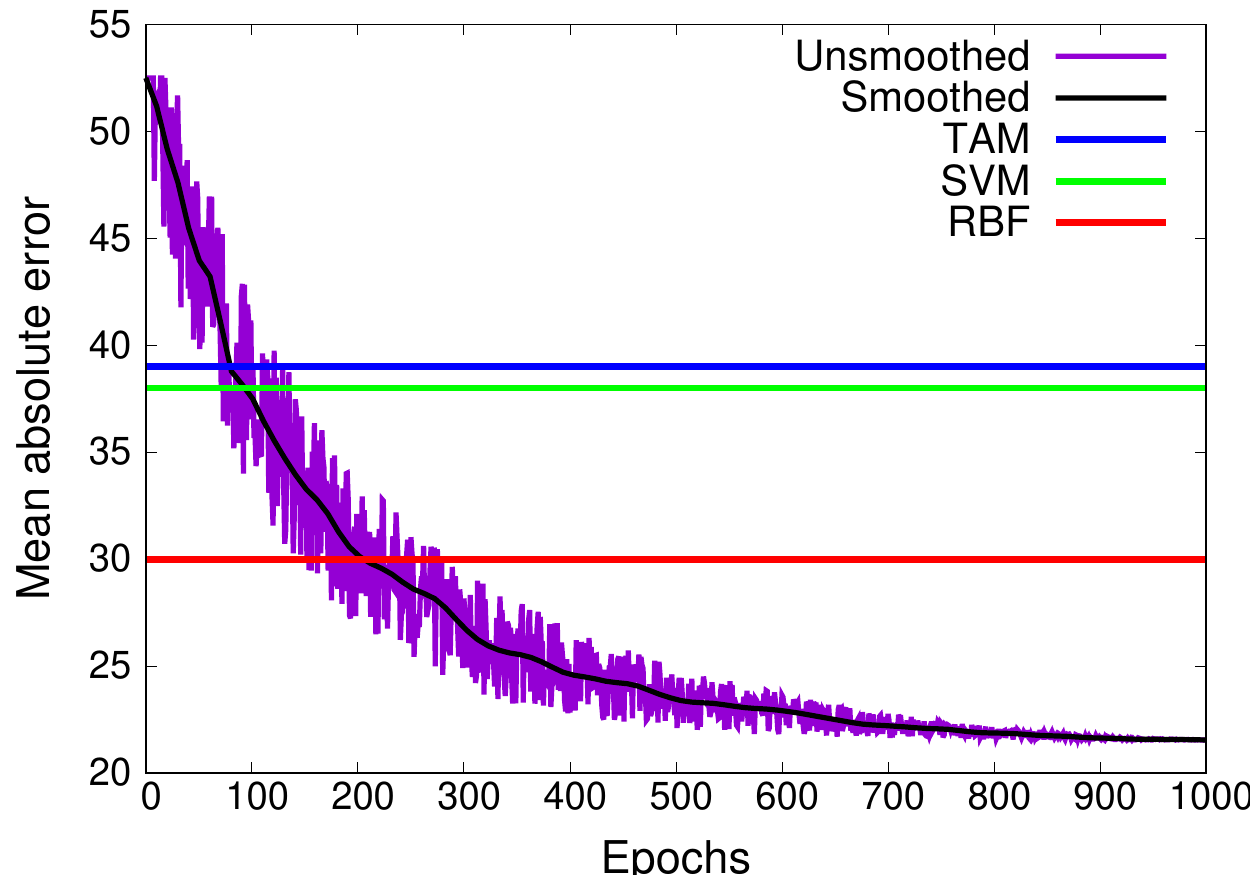}
    \caption{\small Training convergence (TPC-DS)}
    \label{fig:train_ds}
  \end{subfigure}
  \caption{Training overhead results}
\end{figure*}
  
  \sparagraph{Optimizations} Here, we evaluate the two training optimizations introduced in Section~\ref{sec:model-training}: (a) \emph{information sharing}, caching and reusing computations that are shared by multiple nodes, and (b) \emph{batch sampling}, grouping trees with similar structures into batches in order to take advantage of vector processing.

We evaluated these optimizations by training the neural network until the network converged to the best-observed accuracy. We trained the network with no optimizations, the batching optimization alone, the information sharing optimization alone, and with both optimizations. Figure~\ref{fig:train_opts} shows the results for both datasets. Without either optimization, training the neural network for either dataset takes well over a week. Of the new optimizations, information sharing is the more significant in these experiments, bringing the training time down from over a week to a little under 3 days. Both optimizations combined bring the training down to only slightly over 24 hours.

We also measured the memory usage of the information sharing approach, which requires additional space to cache results. The size of the cache was minuscule in comparison to the size of the neural network's weights, with the cache size never exceeding 20MB. We conclude that \emph{both the information sharing and batch sampling optimizations are worthwhile for accelerating the time needed to train the neural network.}

\sparagraph{Training convergence}
Next, we investigate the performance of the model over time during training. After each training epoch (a full pass over the training queries), we evaluated and recorded the mean absolute error across the test set. The results for TPC-H are shown in Figure~\ref{fig:train_h}, and the results for TPC-DS are shown in Figure~\ref{fig:train_ds}. On both plots, the blue, green and red lines show the performance of \texttt{TAM}, \texttt{SVM}, and \texttt{RBF}, respectively. While the neural network model did not converge until epoch 1000 (approximately 28 hours for both datasets), the performance of the neural network begins to exceed the performance of \texttt{SVM} at around epoch 250 (7 hours) with TPC-H, or after 150 epochs (4.5 hours) for TPC-DS. The neural network begins to exceed the performance of \texttt{RBF} after around 350 epochs (10 hours) for TPC-H, or after 250 epochs for TPC-DS (7 hours). \emph{We conclude that the training overhead required for QPP Net to achieve competitive performance is reasonable for a variety of datasets.}

Both Figure~\ref{fig:train_h} and Figure~\ref{fig:train_ds} show the classic, inverse-exponential behavior of neural networks during training. At the start of the training, the neural network decreases its mean absolute error by 20 in just 100 epochs (epochs 0 through 100). However, later in the training, it takes 100 epochs to decrease the mean absolute error by just 2 (epochs 400 through 500). Thus, each additional epoch returns smaller and smaller benefits, requiring a large number of additional epochs to get small gains towards the end of training.

It is possible that using other optimization methods besides stochastic gradient descent, such as Adam~\cite{adam}, might speed up training. We leave such experiments to future work.

\subsection{Network architecture}
For all of the experiments so far, we have used 5 hidden layers each with 128 neurons for each neural unit. While a 5 layer, 128 neuron network would be considered rather small by modern standards, one needs to consider that in our model, \emph{each} neural unit in the plan-based neural network has these dimensions. When assembled together into a tree, the network is much larger (one to two orders of magnitude). However, choosing the number of hidden layers and number of neurons is a difficult task. While there is no theoretically-best number of hidden layers or number of neurons per layer, good values can be found experimentally. Intuitively, ``deeper'' (i.e., more hidden layers) architectures enable more feature engineering, as additional layers add additional transformations of the inputs~\cite{deep_learning}. On the other hand, ``taller'' (i.e., larger hidden layers) architectures allow each feature transformation to be richer, as they have more weights and thus carry more data~\cite{dnn}.

We analyze four of the variables at play when trying to find the correct network configuration: the number of hidden layers, the number of neurons, the maximum accuracy the network can reach, and the time it takes to train the network. Generally, increasing either the number of hidden layers or the number or neurons results in an increase in training time due to an increase in the number of weights.\footnote{Larger networks also take longer to make predictions at inference time, creating potential complications for applications.} Thus, setting the number of neurons or number of hidden layers too high will result in unacceptably long training times. But, if the number of neurons or hidden layers is set too low, the network might not have enough weights to learn the underlying data distribution well enough. We thus seek the number of neurons and hidden layers that will minimize training time while still giving peak or near-peak accuracy. 

\begin{figure}
\centering
\includegraphics[width=0.42\textwidth]{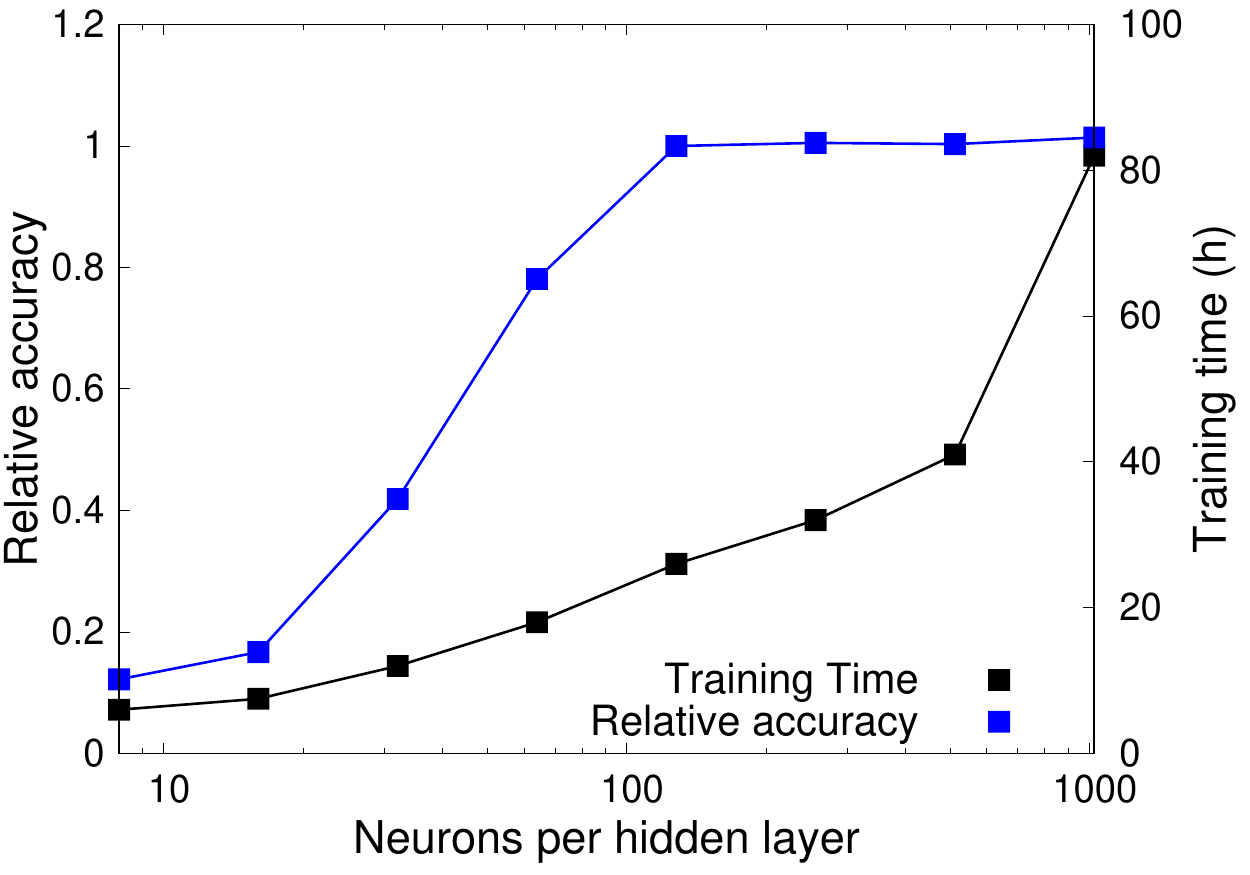}
\caption{ Effect of the number of neurons on accuracy (relative to 128 neurons) and training time.}
\label{fig:neurons}
\vspace{2mm}
\end{figure}

\sparagraph{Varying the number of neurons} Figure~\ref{fig:neurons} shows the training time and relative accuracy  when varying the number of neurons inside each of the five hidden layers. With an extremely small number of neurons (8 neurons), training time is low (6 hours), but accuracy is extremely poor: QPP Net achieves less than 15\% of the accuracy that the 128-neuron network does. On the other hand, using an extremely large number of neurons causes the training time to skyrocket: with 1024 neurons per hidden layer, training time is nearly four times what is required for the 128 neuron network, with only a tiny increase in accuracy (less than 1\%).

One may notice that the training time seems to grow with the log of the number of neurons at first, but then eventually becomes linear. This is because neural networks are trained on GPUs equipped with highly-parallel vector processing units. There is thus sublinear increases in training time until there is approximately one weight per vector processing core, after which the training time changes as expected. When the number of neurons greatly exceeds the capacity of the GPU, the slowdown will become worse than linear.


\begin{figure}
\centering
\includegraphics[width=0.42\textwidth]{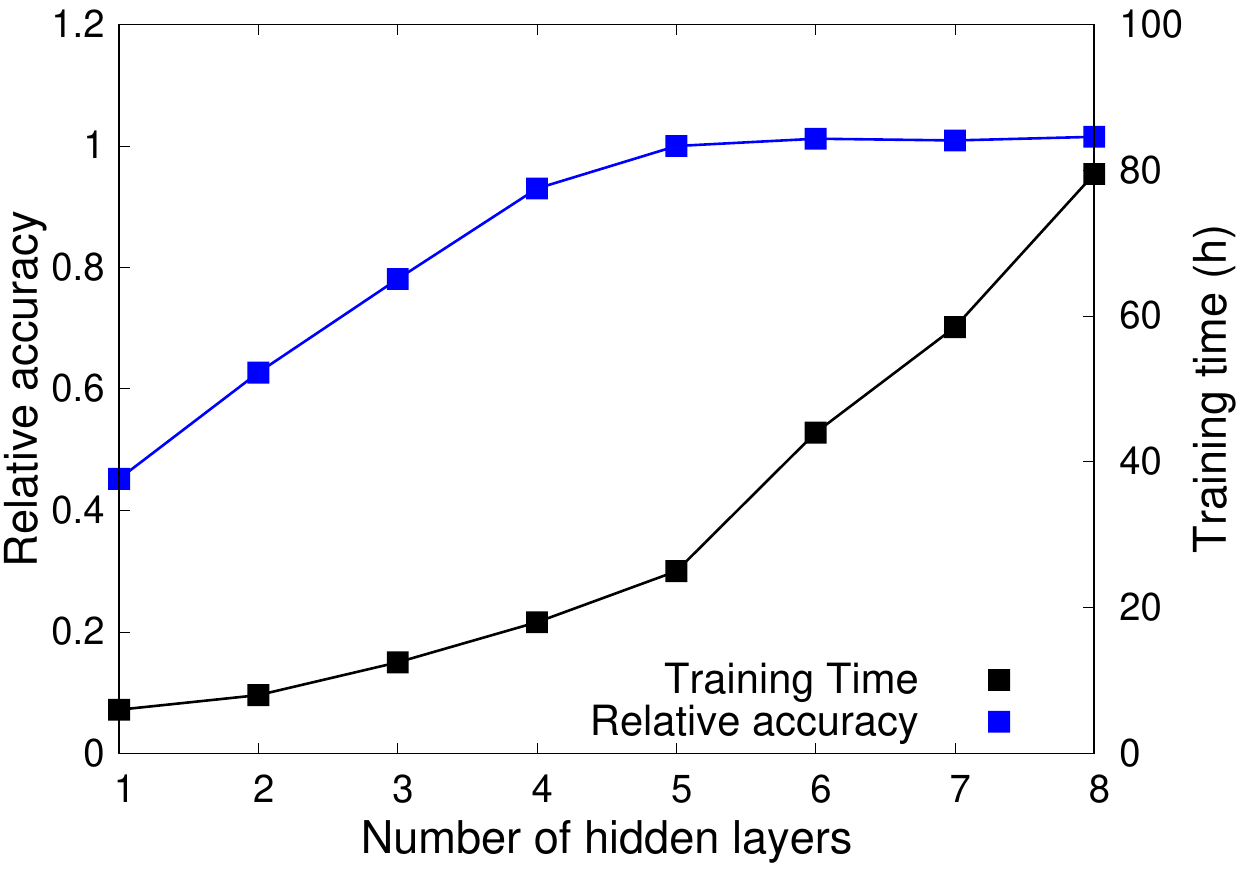}
\caption{ Effect of the number of hidden layers on accuracy (relative to 5 hidden layers) and training time.}
\label{fig:layers}
\end{figure}

\sparagraph{Varying the number of hidden layers}
Figure~\ref{fig:layers} shows a similar experiment, this time varying the number of hidden layers and keeping the number of neurons fixed at 128. Note that connecting two layers with 128 neurons to each other requires a matrix of size $128 \times 128$, so each additional hidden layer adds on the order of $2^{14}$ additional weights.

Increasing the number of hidden layers has a similar behavior to increasing the number of neurons: initially, each increase brings about a small increase in training time but a large jump in accuracy. Eventually, adding another hidden layer produces a much larger jump in training time and a much smaller jump in accuracy. From Figure~\ref{fig:layers}, we can conclude that adding more than 5 hidden layers, at least when the size of each hidden layer is 128 neurons, does not bring about much benefit.


\nextSec
\section{Related work}\label{sec:related}
\sparagraph{Query performance prediction} There exist a number of approaches that leverage machine learning and statistical analysis to address the problem of query performance prediction. We already discussed and compared with~\cite{learning_latency,LiRobustestimationresource2012,opt_est}  in our experimental study.  \cite{pred_multiple} focuses on predicting multiple query  resource usage metrics simultaneously (but not execution times). Both~\cite{pred_xml, pred_xml2} predict statistics about queries in XML databases. \cite{opt_est_par} demonstrating that optimizer cost models can be used to predict query performance if one is willing to sample a percentage of the underlying data.

All these techniques suffer from similar drawbacks: first, they require human experts to analyze the properties of an operator or \QEP and determine how they should be transformed into features for a machine learning algorithm, whereas our deep-learning approach requires no such feature engineering. Second, while some of these approaches model plans, operators, or a combination thereof, none of them learn the \emph{interactions} between various combinations of operators, as the approach presented here does.

A number of techniques~\cite{contender,jennie_sigmod11,ernest} extend to concurrent query performance prediction for analytical queries.  These techniques assume a-priori knowledge of query templates~\cite{jennie_sigmod11},  query structure~\cite{ernest} and/or require extensive offline training on representative queries~\cite{contender}. Furthermore, their proposed input features, metrics and models are hand-tuned to handle only analytical tasks, which make them less applicable to diverse workloads. 


\sparagraph{Cardinality estimation} Works on cardinality estimation are fundamentally related to query performance prediction, as an operator's cardinality often correlates with its latency. Techniques for cardinality estimation include robust statistical techniques~\cite{card_sampling, bound_card, robust_qo}, adaptive histograms~\cite{leo,AboulnagaSelftuningHistogramsBuilding1999}, and deep learning~\cite{deep_card_est,deep_card_est2}. While query cardinalities are certainly an indicator of latency, translating even an accurate cardinality estimate for each operator in a \QEP to a total plan latency is by no means a trivial task. However, a technique predicting operator cardinalities could be easily integrated into our deep neural network by inserting the cardinality estimate of each operator into its neural units input vector. The neural network could then learn the relationship between these estimates and the latency of the entire \QEP.

\sparagraph{Progress estimators} Work on query progress indicators~\cite{MortonParaTimerProgressIndicator2010,LuoProgressIndicatorDatabase2004,LiGSLPICostBasedQuery2012,XiePIGEONProgressindicator2015,LeeOperatorQueryProgress2016} essentially amounts to frequently updating a prediction of a query's latency. These approaches estimate the latency of a query \emph{as it is running}, and the estimate that these techniques make at the very start of the query's execution may be quite inaccurate, but are quickly refined and corrected during the early stages of a query's progress. This greatly limits their applicability for ahead-of-time \QPP, and thus we do not compare against any of these techniques directly.


\sparagraph{Deep learning} We are not the first to apply deep learning to database management problems. Deep learning has seen a recent groundswell of activity in the systems community~\cite{dbml}, including several works on query optimization~\cite{rejoin, qo_state_rep}, entity matching~\cite{deep_entity}, index selection~\cite{lift,selfdrivingcidr}, indexes themselves~\cite{ml_index}, and cardinality estimation~\cite{deep_card_est, deep_card_est2}.

\vspace{1em}
\section{Conclusions and future work}\label{sec:conclusions}
We have introduce a novel neural network architecture designed specifically to address the challenges of query performance prediction. The architecture allows for {plan-structured neural networks} (networks that predict the execution time of a given query plan)  to be constructed by assembling operator-level neural units (neural networks that predict the latency of a given operator) to form a tree-based structure that matches the  structure of the query plan generated by the optimizer. We motivated the need for this novel model, described its architecture and have shown how the model can be effectively trained through two optimizations. Experimental results demonstrate that our approach outperforms state-of-the-art solutions, with manageable training overhead. 


Future work could advance in a number of directions. For example, the neural network architecture presented here could be adapted to handle concurrent queries. Doing so would require understanding the resource usage requirements of the two queries, and whether or not two queries will have to compete for resources. Furthermore, the neural network architecture presented here was used to predict the performance of queries executed on a bare-metal server. However,  in a cloud environment, performance can vary based on the time of day, or seemingly randomly. Techniques to capture the relative performance of the system (e.g., monitoring I/O rates, CPU cycles, etc.), and ways of making the neural network model aware of performance fluctuations, could be investigated.
Improvements to the training time of neural network could be investigated. For example, using a different optimizer~\cite{adam} or taking advantage of transfer learning techniques~\cite{transfer, transfer2} may prove fruitful.


\nextSec

\clearpage
\bibliographystyle{abbrv}
\bibliography{ryan-cites-short}

\appendix

\section{TPC-DS Template Latency}
\label{apx:latency}
\begin{figure*}
\centering
\includegraphics[width=\textwidth]{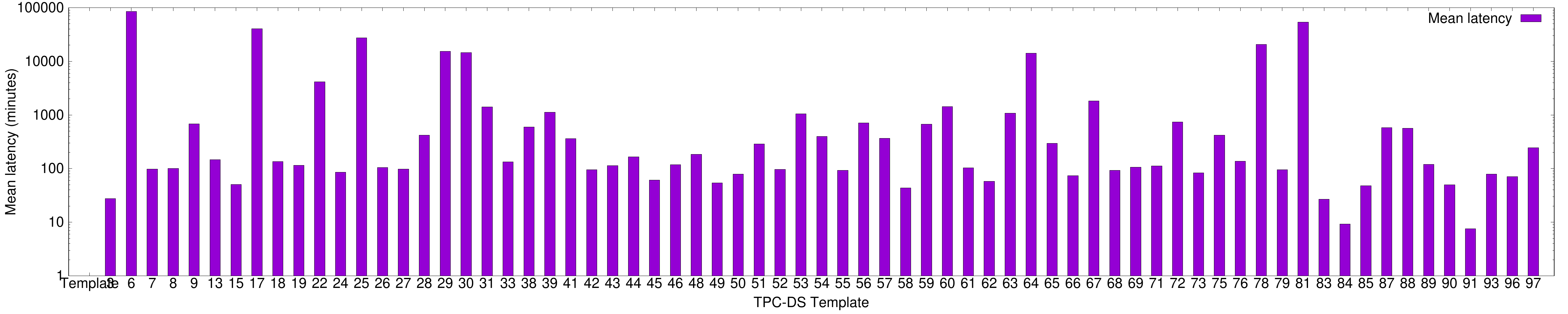}
\caption{\small Mean query latency by template for TPC-DS}
\label{fig:mean_by_template_ds}
\end{figure*}

Figure~\ref{fig:mean_by_template_ds} shows the mean latency for each of the TPC-DS query templates.

\section{Features}\label{apx:features}

\begin{table*}[]
\begin{tabular}{cccl}
\toprule
  Feature             & PostgreSQL operators & Encoding & Description \\ \midrule
  Plan Width          & All                & Numeric  & Optimizer's estimate of the width of each output row \\
  Plan Rows           & All                & Numeric  & Optimizer's estimate of the cardinality of the output of the operator \\
  Plan Buffers        & All                & Numeric  & Optimizer's estimate of the memory requirements of an operator \\
  Estimated I/Os      & All                & Numeric  & Optimizer's estimate of the number of I/Os performed \\
  Total Cost          & All                & Numeric  & Optimizer's cost estimate for this operator, plus the subtree \\ \midrule
  Join Type           & Joins              & One-hot  & One of: semi, inner, anti, full \\ 
    Parent Relationship & Joins              & One-hot  & When the child of a join. One of: inner, outer, subquery \\ 
  Hash Buckets        & Hash               & Numeric  & \# hash buckets for hashing \\ 
  Hash Algorithm      & Hash               & One-hot  & Hashing algorithm used \\
    Sort Key          & Sort                & One-hot  & Key for sort operator\\
  Sort Method         & Sort              & One-hat  & Sorting algorithm, e.g. ``quicksort'', ``top-N heapsort'', ``external sort''\\ \midrule
  Relation Name       & All Scans          & One-hot  & Base relation of the leaf \\
  Attribute Mins      & All Scans          & Numeric  & Vector of minimum values for relevant attributes \\
  Attribute Medians   & All Scans          & Numeric  & Vector of median values for relevant attributes \\
  Attribute Maxs      & All Scans          & Numeric  & Vector of maximum values for relevant attributes \\
  Index Name          & Index Scans         & One-hot  & Name of index \\
  Scan Direction      & Index Scans        & Boolean  & Direction to read the index (forward or backwards) \\ \midrule
  Strategy            & Aggregates         & One-hot  & One of: plain, sorted, hashed \\
  Partial Mode        & Aggregate          & Boolean  & Eligible to participate in  parallel aggregation \\
  Operator            & Aggregate          & One-hot  & The aggregation to perform, e.g. \texttt{max, min, avg} \\
\end{tabular}
\vspace{3mm}
\caption{QPP Net Inputs}
\label{tbl:features}
\end{table*}

Table~\ref{tbl:features} describes the values used as inputs for our neural units. The first column lists the name of the quantity. The second column describes which PostgreSQL operators use a particular type of input. The third column describes how the particular value is encoded into an input suitable for a neural network. The encoding strategies are:
\begin{itemize}
\item{{\bf Numeric}: the value is encoded as a numeric value, scaled so that the mean of the value across the training set is zero and the variance is one. At inference time, the same scaling values are used. This is known as ``whitening'', and is a standard practice in deep learning~\cite{pytorch}.}
\item{{\bf Boolean}: the value is encoded as either a zero or a one.}
\item{{\bf One-hot}: the value is categorical, and is encoded as a one-hot vector, e.g. a vector with a single ``1'' element where the rest of the elements are ``0''.}
\end{itemize}

The first five values, in the first section of the table, represent information that is available for every PostgreSQL operator. Thus, these values are included in all neural units. The next section of the table (``Join Type'' to ``Sort Method''), corresponds to information used by the join neural unit. The third section (``Relation Name'' to ``Scan Direction'') refers to the inputs used for the scan neural unit, and, depending on the selected physical operator type (e.g., index scan, table scan), some values may be missing. Missing values are set to zero. The final section of the table (``Strategy'' to ``Operator'') lists the inputs used for the aggregate neural unit.


\end{document}